\providecommand{\e}[1]{\ensuremath{\times 10^{#1}}}
\documentclass[12pt,twocolumn]{emulateapj}
\usepackage{natbib}
\usepackage{epstopdf}
\usepackage{amsmath}

\begin{document}
\title{A multiple scattering polarized radiative transfer model: application to HD 189733\MakeLowercase{b} }
\author{Pushkar Kopparla\altaffilmark{1,*}, Vijay Natraj\altaffilmark{2}, Xi Zhang\altaffilmark{3,4} Mark R. Swain\altaffilmark{2}, Sloane J. Wiktorowicz\altaffilmark{5}, Yuk L. Yung\altaffilmark{1}}
\altaffiltext{1}{Division of Geological and Planetary Sciences, California Institute of Technology, Pasadena, CA}
\altaffiltext{2}{Jet Propulsion Laboratory (NASA-JPL), Pasadena, CA}
\altaffiltext{3}{Lunar and Planetary Laboratory, University of Arizona, Tucson, AZ}
\altaffiltext{4}{Earth and Planetary Sciences Department, University of California, Santa Cruz, CA}
\altaffiltext{5}{Department of Astronomy and Astrophysics, University of California, Santa Cruz, CA}
\altaffiltext{*}{Corresponding author, email:pkk@gps.caltech.edu}
\begin{abstract}
{We present a multiple scattering vector radiative transfer model which produces disk integrated, full phase polarized light curves for reflected light from an exoplanetary atmosphere. We validate our model against results from published analytical and computational models and discuss a small number of cases relevant to the existing and possible near-future observations of the exoplanet HD 189733b. HD 189733b is arguably the most well observed exoplanet to date and the only exoplanet to be observed in polarized light, yet it is debated if the planet's atmosphere is cloudy or clear. We model reflected light from clear atmospheres with Rayleigh scattering, and cloudy or hazy atmospheres with Mie and fractal aggregate particles. We show that clear and cloudy atmospheres have large differences in polarized light as compared to simple flux measurements, though existing observations are insufficient to make this distinction. Futhermore, we show that atmospheres that are spatially inhomogeneous, such as being partially covered by clouds or hazes, exhibit larger contrasts in polarized light when compared to clear atmospheres. This effect can potentially be used to identify patchy clouds in exoplanets. Given a set of full phase polarimetric measurements, this model can constrain the geometric albedo, properties of scattering particles in the atmosphere and the longitude of the ascending node of the orbit. The model is used to interpret new polarimetric observations of HD 189733b in a companion paper.}

\end{abstract}
\section{Introduction}

Polarimetry has been used to probe the atmospheres of planets in the solar system; the first observation of linear polarization from Venus' atmosphere is credited to \citet{lyot1929pol}.  More recently, the properties of clouds and hazes in Venus's atmosphere were deduced through polarimetric data from both ground-based observations and from Pioneer data \citep{hansen1974interpretation,kawabata1980cloud}.  Similar successful studies exist for Titan \citep{tomasko1982photometry}, Jupiter \citep{smith1984photometry} and the other outer planets \citep{joos2007polarimetry}. The idea of using polarimetry to probe exoplanetary atmospheres is thus a natural extension, and was first examined in a theoretical study by \citet{seager2000photometric}. The great advantage of using polarimetry in the study of exoplanets is the increase in contrast between direct starlight and the reflected light from the planetary atmosphere. Integrated over the whole disk, direct starlight from inactive, nearby stars can be assumed to be unpolarized to a high degree\footnote{Polarization is introduced in starlight through interactions with interstellar dust clouds and magnetic fields therein \citep{leverett1949polarization}. However these values are unlikely to vary on planetary orbital timescales.}. For instance, the linear polarization integrated over the sun's disk is $\sim 1 ppm$ (parts per million) in visible wavelengths \citep{kemp1987optical}, light scattered from a planetary atmosphere may have polarizations of a few tens of percent.  Thus, depending on the reflectivity of a planetary atmosphere, the degree of polarization in the star-planet system can be dominated by the reflected light from the planetary atmosphere. In such a case, the combined star planet system should show a periodic modulation in the degree of polarization as the planet moves through different phases of illumination in its orbit. 

A prime exoplanet candidate for polarimetric studies is HD 189733b, a hot Jupiter orbiting a K star, with a semimajor axis of 0.031 AU. The system is relatively close by (19.3 parsecs) and thus bright. \citet{berdyugina2008first} reported a detection of polarized light of amplitude $200 ppm$. Surprisingly, the strength of observed polarization was about one order of magnitude higher than predicted assuming a semi-infinite Rayleigh scattering atmosphere (which produces the highest degree of polarization for a given planetary radius), leading to some skepticism over the observations \citep{lucas2009planetpol}. Follow up studies since  have not reached a consensus on the observed degree of polarization \citep{wiktorowicz2009nondetection, berdyugina2011polarized}. Furthermore, \citet{lucas2009planetpol} observed the polarization of two other exoplanet systems, 55 Cnc and $\tau$ Boo. In both cases, they found polarization of the order of $1 ppm$ but there was no significant variability associated with the orbital periods of the known exoplanets in these systems. In parallel however, there have been few efforts to model the observable polarization signal using what is known about the atmosphere of HD189733b from photometric measurements, since the early work of \cite{lucas2009planetpol} and \cite{sengupta2008cloudy}. We will briefly examine the observations, and some of the issues involved in their interpretation, in order to understand what information can be retrieved using a multiple scattering radiative transfer model.

The paper is structured as follows. The remainder of the introduction is devoted to a review of exoplanetary polarization studies, both theoretical and observational. In Section 2, we outline our model setup and validate our model using observations of Jupiter. In Section 3, we discuss the observable polarization signal for different atmospheric compositions, orbital orientations and spatial inhomogeneities in the atmosphere, followed by a summary in Section 4.
\subsection{Polarimetry of HD 189733b}

\citet{berdyugina2008first}'s study consisted of 93 individual nightly observations taken in the B band (370-550nm) through the KVA 0.6-m telescope and find variable polarization of amplitude $200ppm$. {The degree of polarization is always measured as a fraction of the direct starlight, and not just the reflected light from the planet.} They interpret their observations using a single scattering Rayleigh-Lambertian model, and are able to retrieve values of eccentricity and orbital inclination that agree quite well with other studies. To explain the large degree of polarization, they are forced to use a large planetary radius, 1.5 $\pm$ 0.2 $R_J$ where the standard value is 1.154 $\pm$ 0.017 $R_J$ \citep{pont2007hubble}. They comment that this large radius might be indicative of an extended, evaporating halo around the planet. It is uncertain if such a halo would be reflective enough to be responsible for a significant fraction of the reflected intensity. 

\citet{wiktorowicz2009nondetection} observed the same planet in the wavelength range 400-675 nm from the Palomar 5-m telescope.  He found polarization of the order of $10 ppm$, but there was no significant relationship with the period of the exoplanet. However, this study has only one observation near elongation (phase angle 90$^{\circ}$) where polarization is expected to peak and most observations are at phases where polarization is expected to be small, as has been pointed out by later papers \citep{berdyugina2011polarized}. This study also derives an upper limit to the polarimetric modulation of the exoplanet as $79 ppm$ and the polarimetric variability of starspots to $21 ppm$.

\citet{berdyugina2011polarized} observed polarization modulations in three different bands U, B, V bands centered at 360, 440 and 530 nm respectively with the NOT 2.5-m telescope. They find that the degree of polarization varies with the wavelength in proportion to the Rayleigh scattering law across the different bands. However, they revise their earlier value of the amplitude of polarization from $200 ppm$ down to $\sim 100 ppm$.  Because of the visual similarities to Neptune in the geometric albedo profile, they suggest that the atmosphere might have a similar structure, with a high altitude haze layer above a semi-infinite cloud deck. Another proposed structure is the presence of a dust condensate layer beneath a thin gas layer. 

\subsection{Photometric Observations}
Temperatures in the atmosphere of HD 189733b{are thought to} vary between 1000-1500 K depending on altitude and longitude \citep{knutson2009multiwavelength,knutson2012phase,huitson2012temperature}.  {Its atmosphere is fairly well studied and is known to contain water \citep{tinetti2007water}, carbon monoxide \citep{de2013detection}, carbon dioxide and methane \citep{swain2008presence,swain2009water} in trace amounts. The bulk composition is usually modeled to be mostly hydrogen and helium \citep{huitson2012temperature,danielski20140}.} From theoretical models, it is also expected that such an atmosphere would contain traces of metals like sodium, potassium and magnesium \citep{fortney2010transmission}. Weak detections of these metals from visible \citep{redfield2008sodium} and infrared transmission spectra as well as strong slope from the UV to the near infrared, lead to the inference that a high level, Rayleigh haze that spans several scale heights over an opaque cloud deck may be present \citep{sing2009transit,desert2011transit,sing2011hubble,pont2013prevalence}.   

However, a recent pair of studies \citep{crouzet2014water,mccullough2014water} have put forth an alternative interpretation of the transit and secondary eclipse data. They argue that the slope previously attributed to a Rayleigh scattering haze could instead be caused by unocculted star spots in the field of view. This interpretation favors a clear, cloudless atmosphere for HD 189733b, though it does not rule out a hazy atmosphere.

The geometric albedo of HD 189733b was measured by \citet{evans2013deep} using the HST to measure the brightness of the disk at secondary eclipse, and they find values of $0.40\pm 0.12$ in the range 290-450 nm and an upper limit of 0.12 between 450-570 nm. This data provides an independent check for the albedos retrieved by \citet{berdyugina2011polarized}. The values of  \citet{berdyugina2011polarized} are systematically higher than those obtained by \citet{evans2013deep}.

\subsection{Theoretical Polarization Studies}
\citet{seager2000photometric} in a pioneering study produced theoretical polarization curves for reflected planetary light using a {forward} Monte Carlo radiative transfer (RT) model for a Rayleigh scattering atmosphere. They concluded that the maximum degree of polarization (1-5\e{-5}) was in most cases below detection limits at the time. They also examined the effects of scattering particles and cloud layers, in all cases deviation from a purely scattering gaseous atmosphere reduced the degree of polarization. Following this, there were a series of papers e.g., \citet{stam2004using,stam2006integrating} using an adding doubling RT model. Their results were similar to that of the previous work, but the great advantage of their model is the generation of a "planetary scattering matrix" using which a single calculation can replicate multiple scattering radiative transfer through a planetary atmosphere of arbitrary thickness and composition (only for top of the atmosphere fluxes). \citet{buenzli2009grid} explored the dependence of observable polarization signals on single scattering albedo, optical depth of the scattering layer, and albedo of an underlying Lambert surface for purely Rayleigh scattering atmospheres using a Monte Carlo model. \citet{madhusudhan2012analytic} used an analytic model on a Rayleigh scattering atmosphere to map out polarization signals for various scenarios.

{While extensive parameter searches have been performed theoretically, the observational data have only been interpreted using very simple single scattering Rayleigh-Lambert models \citep{berdyugina2008first,berdyugina2011polarized}. 
Newer theoretical studies have moved onto modeling increasingly specialized features such as rainbows from water clouds \citep{bailey2007rainbows,karalidi2013flux}, surface vegetation\citep{stam2008spectropolarimetric}, oceans \citep{williams2008detecting} and relatively fine cloud structure \citep{karalidi2012modeled}. Current observations exist only for hot Jupiters, and it is unlikely that most of these features will either exist or be observable on such planets in the near future. In this way, there is a divergence between the modeling and observational community within the field. The goal of this work is the creation of an atmospheric polarization model with sufficient physics (multiple scattering, use of non-Rayleigh scattering functions, multiple atmospheric layers, inhomogeneous atmospheres) but simple enough (sufficiently few parameters) to be useful in the interpretation of current and near-future observations (so that data can constrain model parameters). The purpose of this model is to be a tool that augments our understanding of observations; it is not intended to function as a stand alone parameter search engine.}

\section{The Atmospheric Structure of HD 189733b and Radiative Transfer}

\subsection{Model Setup}
Our approach to building an exoplanetary atmospheric polarization model is to start with a well understood planetary atmospheric model, continually modify into an atmospheric structure relevant to HD 189733b and validate it at each step. We begin with a model of Jupiter's stratosphere based on retrievals of Cassini data \citep{zhang2013stratospheric}, henceforth Z13. This model is attractive as a baseline since the atmosphere has realistic clouds and two different types of haze particles: spherical and fractal aggregates. While current polarimetric observations may not have sufficient data to distinguish between these two types of haze particles, we are optimistic about the future. Z13 model the atmosphere of Jupiter using a 12-layer plane parallel atmosphere with scattering and absorption at each layer, underlain by a reflective semi-infinite cloud layer. This model currently works only with the photometric intensity, \textit{I}, while we require at least three of the Stokes parameters, \textit{I}, the intensity, and \textit{Q} and \textit{U}, the linear polarization parameters.{The degree of polarization, \textit{P} is defined as
\begin{equation}
P = \frac{\sqrt{Q^2+U^2}}{I_{star}+I_{planet}} \sim \frac{\sqrt{Q^2+U^2}}{I_{star}}
\end{equation}  
We talk about degree of polarization in the total star-planet flux since this is the observable quantity. Direct starlight is assumed to be unpolarized integrated over the disk of the star.} Symmetry breaking due to starspots or the transiting planet itself can induce non-zero polarization in direct starlight. However these effects have been calculated to be about one order of magnitude smaller than the expected polarization from the planet \citep{kostogryz2015polarization}. If the detected polarization of the planet is much smaller than expected, these effects will become important and must be accounted for. The first change we make is to swap out the scalar RT model DISORT \citep{stamnes2000disort} which works only on intensities, with a vector RT model VLIDORT \citep{spurr2006vlidort} that can handle polarized radiances.{ This is a plane-parallel scattering code that uses the discrete ordinate method to approximate multiple scatter integral source terms in the RT equation. The model will make a precise single scatter calculation for both incoming solar and outgoing line-of-sight beams in a plane-parallel or spherical-shell atmosphere. Stokes vector output may be generated at any level in the atmosphere and for any angular distribution, using the source function integration technique. The model can handle coupled thermal/surface emission and multiple scattering scenarios, and there is a provision for dealing with bidirectional reflecting surfaces as well as the usual Lambertian surface assumption. The VLIDORT  model is also fully linearized: simultaneously with the polarized radiance field, it will deliver analytic Jacobians with respect to any atmospheric and/or surface properties.}

{VLIDORT has been validated against Rayleigh \citep{coulson1960tables} and aerosol benchmark results \citep{garcia1989fn,siewert2000discrete}. Details of the validation can be obtained from \citet{spurr2006vlidort}. VLIDORT has also been validated in the thermal infrared (with no solar sources) and mid infrared (with both solar and thermal emission sources) spectral regions by comparisons with the National Center for Atmospheric Research GENLN Spectral Mapper model, which in turn is based on the GENLN line-by-line RT algorithm \citep{edwards1992genln2}.} VLIDORT has been previously used in remote sensing applications for Earth \citep{cuesta2013satellite,xi2015simulated}. 

\begin{figure}[h!]
  \centering

    \includegraphics[width=8cm]{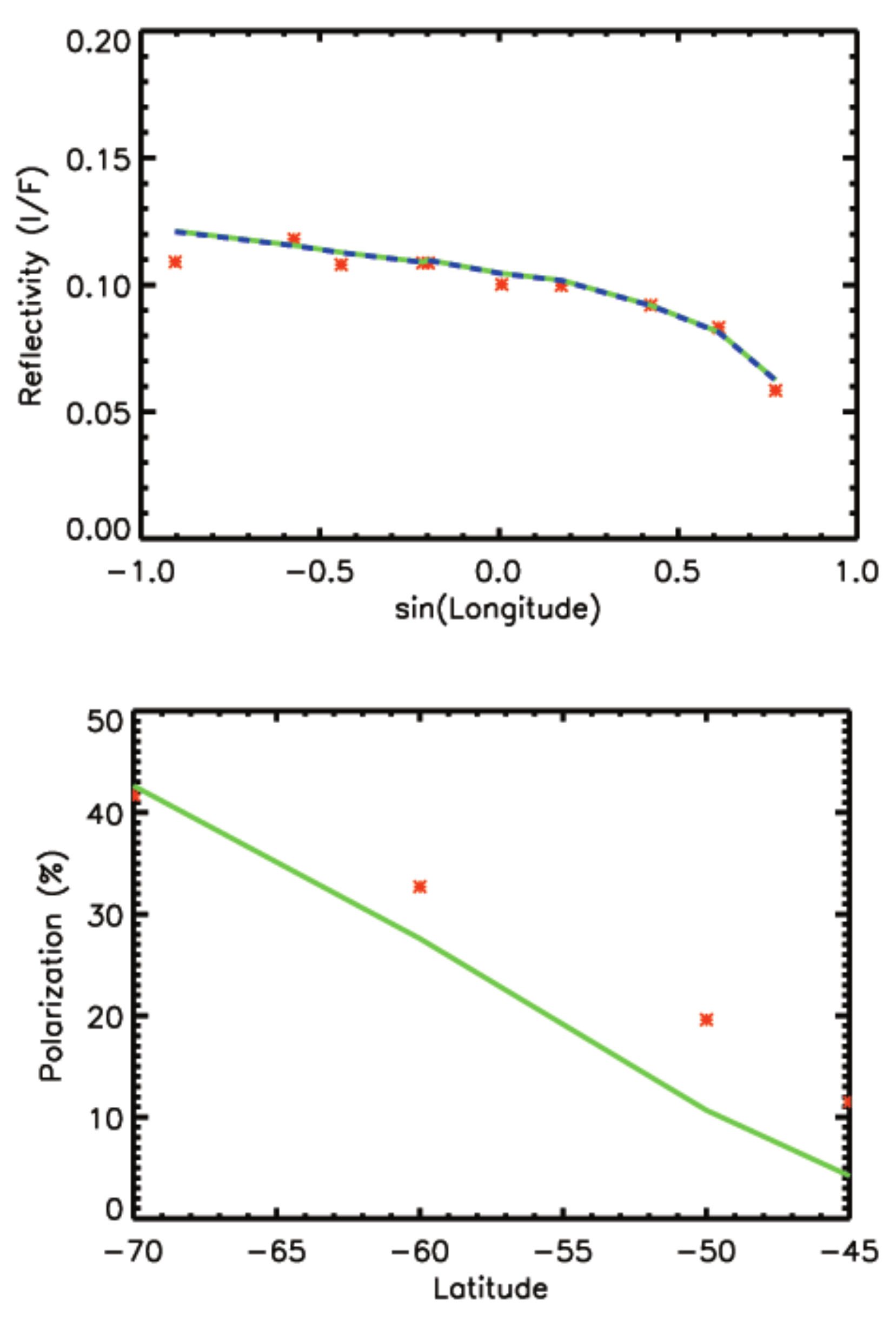}
      \caption{Top panel shows reflectivities from \citet{zhang2013stratospheric} using DISORT(blue dashed line), this work using VLIDORT (solid green line) and observations from Cassini in the UV1 filter (red points) at S60$^{\circ}$ latitude and a phase angle of 17.5$^{\circ}$. This is the blue curve in the top left panel of Figure 8 in \citet{zhang2013stratospheric}. Bottom panel shows observed values of polarization in the blue channel from Pioneer 10  \citep[red points]{smith1984photometry} and the corresponding modeled values (this work, solid green line) at a phase angle of 98$^{\circ}$.}
         \label{fig:jupiterzhang}
\end{figure}

We validate this modified Jupiter atmosphere vector model by reproducing known photometric and polarimetric results for the atmosphere of Jupiter. The basic calculation here is one dimensional, monochromatic radiative transfer in a plane parallel atmosphere for a given set of observing angles using an 8-stream RT model, following Z13.{ We are able to reproduce, up to four decimal places, the best fits of Z13 to Cassini data at different wavelength filters and latitudes. A representative plot is shown in the top panel of Figure 1 for an atmosphere containing fractal aggregate hazes over a reflective cloud layer. Haze particles are either modeled as Mie spheres \citep{de1984expansion} or fractal aggregates, using the approximate method of \citet{tomasko2008model} designed for Titan hazes.} Well resolved polarimetric data of Jupiter has existed since the Pioneer missions \citep{smith1984photometry}. We attempt to reproduce these values using the atmospheric model of Z13 with VLIDORT for the relevant latitudes. The bottom panel of Figure \ref{fig:jupiterzhang} shows the degree of polarization produced by our model and the corresponding Pioneer observations, taken from Table IVC of \citet{smith1984photometry}. These observations correspond to the blue filter, whose central wavelength is $0.44 \mu m$ \citep{pellicori1973pioneer}. The fit is certainly not as good as that for photometry. Inaccuracies may partially be due to the fact that the retrieved parameters are optimized to match the photometric data from Cassini alone, whose wavelengths are different. The optical properties at $0.44 \mu m$ are thus interpolated values from the Cassini retrievals. Also, aerosol properties at $98^{\circ}$ phase angle are not constrained well by the Cassini ISS data in Z13, most of the images of which are at small ($<3 0^{\circ}$) and large ($> 130^{\circ}$) phase angles. {Two more validation cases are discussed in the following section on disk integration.}

\subsection{Disk Integration}

Unlike Jupiter, where the disk of the planet is well resolved, polarimetric observations of exoplanets will only yield disk integrated values. Thus, the equation of radiative transfer must be solved for a finite number of points on the planet and the emergent radiance summed to yield disk integrated values. We use the quadrature method of \citet{horak1950diffuse} for fast disk integration.{The polarized light can be represented by the Stokes vector, \textbf{I}, which comprises of the four Stokes parameters, \textit{I}, \textit{Q}, \textit{U} and \textit{V}.
\begin{equation}
\textbf{I} = \begin{bmatrix}
I \\
Q \\
U \\
V \\
\end{bmatrix}
\end{equation}
}
The integral of interest, which gives the integrated Stokes parameters of the planet over the illuminated fraction of the disk at a phase angle $\alpha$ is 
\begin{equation}
j(\alpha) = \int_0^\pi \! sin^2\eta \ d\eta \int_{\alpha-\pi/2}^\pi \textbf{I}(\eta,\zeta) cos\zeta d\zeta 
\end{equation}

Where $\textbf{I}(\eta,\zeta)$ is the outgoing Stokes vector from the point defined by the colatitude $\eta$ and longitude $\zeta$ in the direction of the observer. The intensity within the integral is not analytical and must be obtained from multiple scattering calculations from VLIDORT. It is therefore preferable to have the integral expressed as a summation over some {finite} number of points. Using the transformation
\begin{equation}
\xi = \left(\frac{2}{cos\alpha + 1}\right)\nu + \left(\frac{cos\alpha -1}{cos\alpha +1}\right), \\
\psi = cos\eta
\end{equation}
where $\nu = \sin \eta$, the limits of the integrals are changed to -1 to +1. {Note that these equations are valid for positive phase angles, $\alpha$. For negative phase angles, the extent of the illuminated disk is expressed as
\begin{equation}
j(\alpha) = \int_0^\pi \! sin^2\eta \ d\eta \int_{-\pi/2}^{\pi-\alpha} \textbf{I}(\eta,\zeta) cos\zeta d\zeta 
\end{equation}
 The corresponding variable substitution is now  
\begin{equation}
\xi = \left(\frac{2}{cos\alpha + 1}\right)\nu - \left(\frac{cos\alpha -1}{cos\alpha +1}\right), \\
\psi = cos\eta
\end{equation}}

These integrals can now be expressed as the summations
\begin{equation}
j(\alpha )=\frac{(cos\alpha +1)}{2}\sum \limits_{i=1}^n \sum \limits_{j=1}^n w_i u_j \textbf{I}(\psi _i,\xi _j)
\end{equation}
where each $w_i$ and $u_j$ represents the quadrature weights for the quadrature divisions $\psi_i$ and $\xi_j$. For a given number of summation terms, n, the quadrature weights and divisions are well defined \citep{chandrasekhar1960radiative}. VLIDORT only needs to be run at the positions on the disk indicated by these divisions and summed up to give the outgoing intensity. {The inputs to VLIDORT are the solar zenith angle ($\theta _o$), indicating the direction of the incoming flux from the star measured with respect to the local normal to the surface, viewing zenith angle ($\theta _o$), which is the direction of outgoing radiance to the observer, and the relative azimuthal angle ($\Delta \phi$) between these two directions. These angles are given by 
\begin{equation}
\cos\theta _o  = \sin\eta \cos(\eta - \alpha)
\end{equation}
\begin{equation}
\cos\theta   = \sin\eta \cos(\eta)
\end{equation}
\begin{equation}
\tan\Delta\phi = \frac{\sin\alpha\cos\eta}{\cos\theta\cos\theta _o - \cos\alpha}
\end{equation}}
We verify that our numerical implementation is correct by reproducing Table 3 of \citet{horak1950diffuse} {for surface reflection from a Lambertian surface}.{The effects using different numbers of quadrature points, computational streams in the RT model and the resolution of the orbit are discussed in the appendix. In brief, pure Rayleigh scattering atmospheres are insensitive to resolution effects and use 8-stream, 64-point quadrature. Mie scattering atmospheres require at least 16-stream RT to produce rainbows and use 32-stream and 256-point quadrature for the cases discussed. For inhomogeneous hazy atmospheres with sharp discontinuities in the scattering properties across the disk, 32-stream, 1024-point quadrature was used to produce smooth curves in reflected intensity. Model runtime scales linearly with number of phases modeled per orbit, as the square of the linear spatial resolution of the disk and the cube of the number of RT streams. }

{To further validate our disk integration scheme implementation, we reproduce the disk integrated reflectivity and degree of polarization calculations from \citet{buenzli2009grid} for Rayleigh scattering atmosphere of optical depth 30 and single scattering albedo 0.999999 over a Lambertian surface of albedo 1. Our results agree well with published values, as shown in Table \ref{tab:bstable}. The error in the degree of polarization is given as $0.1\%$ in \cite{buenzli2009grid}. The columns are the scattering phase angle ($\alpha$), reflectivity ($I$, this work, $I_{BS}$ from \citep{buenzli2009grid}) and the degree of polarization expressed as a percentage of reflected light ($q$ and $q_{BS}$ respectively). In this work, we use the definitions of the Stokes parameters as given by \cite{hovenier2014transfer}, which are the same as those used in \cite{chandrasekhar1960radiative}. \cite{buenzli2009grid} use the definition from \cite{coulson1960tables}, which only differs in the sign of \textit{Q}.} 
\begin{table}[h!]
  \begin{center}
    \caption{Comparisons with Rayleigh Scattering Results of Buenzli and Schmid (2009) }
    \label{tab:bstable}
    \begin{tabular}{ccccc}
 \hline
 \hline
      $\alpha[^{\circ}]$ & $I(\alpha)$ &  $I_{BS}(\alpha)$ & $q(\alpha)[\%]$ & $q_{BS}(\alpha)[\%]$  \\
 \hline

2.5 & 0.796 &  0.795 & -0.0 & 0.0\\
7.5 & 0.786 & 0.785 & -0.4  & 0.4\\
12.5 & 0.767 & 0.766 & -1.1 & 1.1\\
17.5 & 0.741 & 0.740 & -2.1 & 2.1\\ 
22.5 & 0.709 & 0.708 &-3.4 & 3.4\\ 
27.5 & 0.672 & 0.671 & -5.0 & 5.1\\ 
32.5 & 0.631 & 0.630 & -6.9 & 6.9 \\
37.5 & 0.588 & 0.587 & -9.1 & 9.1\\ 
42.5 & 0.543 & 0.542 & -11.4 & 11.4\\ 
47.5 & 0.498 &  0.497 & -14.0 & 13.9\\ 
52.5 & 0.453 &  0.453 & -16.6 & 16.6\\ 
57.5 & 0.410 & 0.410 & -19.3 & 19.3\\ 
62.5 & 0.368 & 0.368 & -22.0 & 22.0 \\ 
67.5 & 0.329 & 0.329 &  -24.6 & 24.6\\ 
72.5 & 0.293 & 0.292 & -27.0 & 27.0\\ 
77.5 & 0.259 & 0.259 & -29.1 & 29.1\\ 
82.5 & 0.228 & 0.228 & -30.8 & 30.7\\
87.5 & 0.199 & 0.199 & -32.0 & 31.9 \\ 
92.5 & 0.174 & 0.174 & -32.6 & 32.5\\ 
97.5 & 0.151 & 0.150 & -32.5 & 32.5 \\
102.5 & 0.130 & 0.130 & -31.9 & 31.8\\
107.5 & 0.111 & 0.111 & -30.5 & 30.5\\ 
112.5 & 0.094 & 0.094 & -28.6 & 28.6\\ 
117.5 & 0.079 & 0.079 & -26.2 & 26.2\\ 
122.5 & 0.066 & 0.066 & -23.4 & 23.4\\
127.5 & 0.054 & 0.054 & -20.2 & 20.3\\ 
132.5 & 0.043 & 0.043 & -16.9 & 17.0\\ 
137.5 & 0.033 &  0.033 & -13.6 & 13.7\\
142.5 & 0.025 & 0.025 &-10.3 & 10.4\\ 
147.5 & 0.018 & 0.018 & -7.1 & 7.3\\
152.5 & 0.012 & 0.013 &-4.3 & 4.4\\ 
157.5 & 0.008 & 0.008 & -1.9 & 2.0\\ 
162.5 & 0.004 & 0.005 & 0.1  & 0.0 \\
167.5 & 0.002 & 0.002 &1.5  & -1.4\\
172.5 & 0.001 & 0.001 & 2.0  & -1.9\\
177.5 & 0.000 & 0.000& -  & -\\
 \hline  
    \end{tabular}
     \end{center}
{   Signs are opposite due to the use of different conventions, see text for details }

\end{table}

\begin{figure}[h!]
  \centering

    \includegraphics[width=8cm]{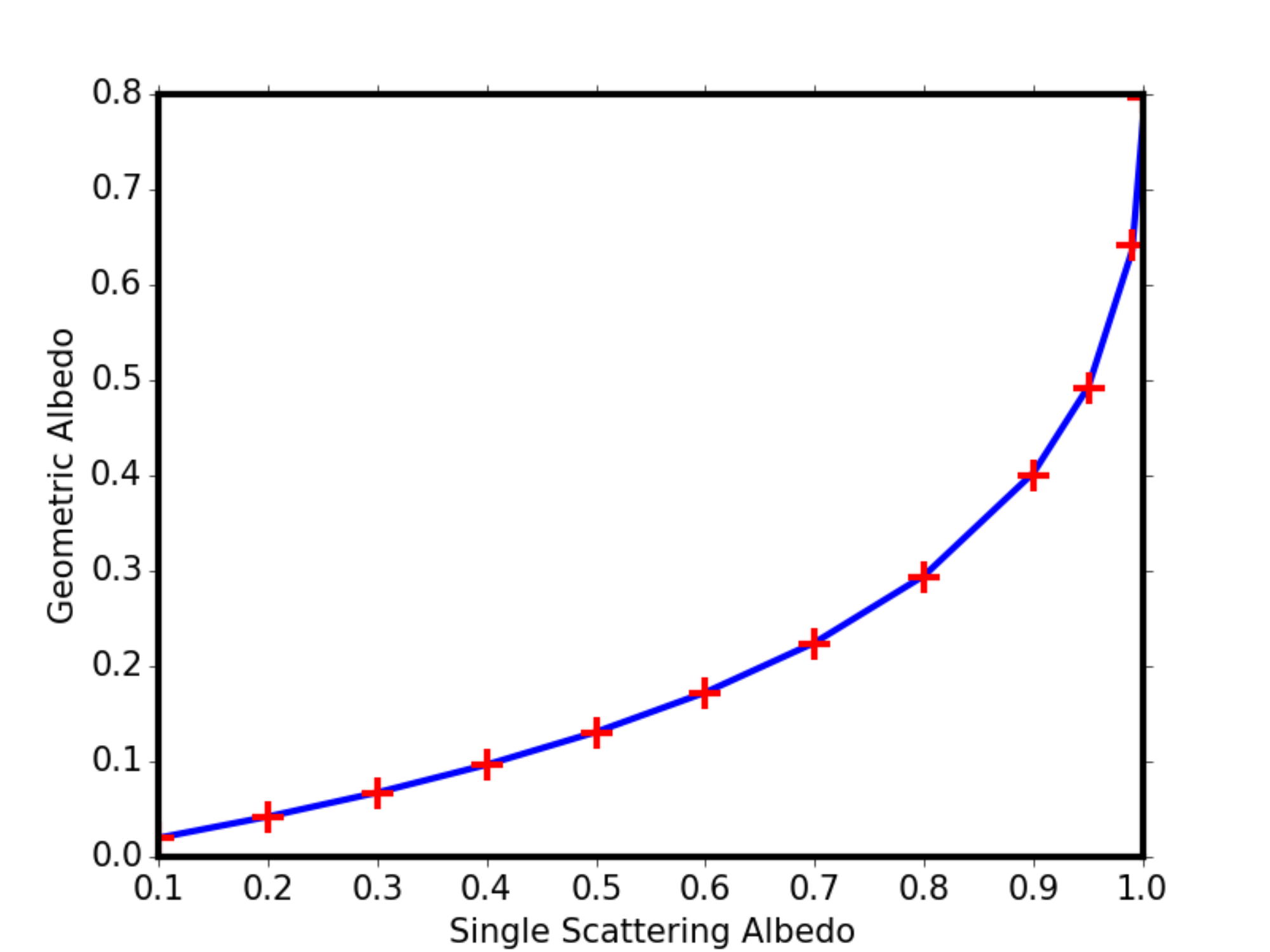}
      \caption{Geometric albedo as a function of single scattering albedo in a semi-infinite Rayleigh scattering atmosphere from this model (blue curve) and from \cite{madhusudhan2012analytic} (red pluses).}
         \label{fig:comparefig5}
\end{figure}

{Furthermore, since we are particularly interested in measurements of the geometric albedo, we also validate our calculations for the dependence of geometric albedo on the single scattering albedo. For comparison, we use the fitted analytic expression from \citet{madhusudhan2012analytic}, shown in Figure \ref{fig:comparefig5}. We get good agreement except close to single scattering albedo $\sim 1$, where the {analytic} fitting expression is not good as reported by \cite{madhusudhan2012analytic}. However, our model value of 0.7976 is close to the published {numerically computed} values of 0.7977 \citep{madhusudhan2012analytic} and  0.7975 \citep{prather1974solution}. VLIDORT cannot handle a single scattering albedo of exactly 1, therefore we use the value 0.999999.}
\subsection{Geometric Considerations}
For a circular orbit, we have the scattering angle for a given orbital position \citep{madhusudhan2012analytic}
 \begin{equation}
 \cos \alpha = \sin \phi \sin i
 \end{equation}
 where $\phi$ is the true anomaly and $i$ is the inclination. As the orbit is circular, we take the true anomaly to be the same as our orbital position angle. $\phi = 0$ corresponds to superior conjunction as seen from Earth and $\phi = \pi$ corresponds to mid-transit. 
The definition of the stokes parameters \textit{Q} and \textit{U} is generally with respect to the plane of the sky as seen from Earth. However, each VLIDORT calculation yields these parameters in the local scattering plane. Therefore, one rotation is necessary  change the reference plane to the equatorial plane of the exoplanet before being summed up by the quadrature formula above. A second rotation is necessary to express the polarizations in the sky plane of the Earth. In both cases, the Stokes parameters \textit{I} and \textit{V} remain unchanged since they deal with the total intensity and the handedness and magnitude of circular polarization. Thus the first rotation is of the form \citep{madhusudhan2012analytic}

\begin{equation}
\begin{bmatrix}
I' \\
Q' \\
U' \\
V' \\
\end{bmatrix}
= \begin{bmatrix}
1 & 0 & 0 & 0 \\
0 & \cos 2 \gamma _1 & \sin 2 \gamma _1 & 0 \\
0 & -\sin 2 \gamma _1& \cos 2 \gamma _1& 0 \\
0 & 0 & 0 & 1 \\
\end{bmatrix} \begin{bmatrix}
I \\
Q \\
U \\
V \\
\end{bmatrix} 
\end{equation}
\begin{equation}
\cos \gamma _1  = \frac{\sin \eta  \ \sin \zeta }{\sin\theta }
\end{equation}
$\theta$ is the angle with the vertical at the point of scattering made by the outgoing beam of radiation to Earth. The angle of the second rotation is a function of the planet's position in the orbit and is given by \citep{schmid1992montecarlo}
\begin{equation}
\gamma _2 = tan^{-1}\left(\frac{\tan\phi}{\cos i}\right)+90^{\circ}+\omega _p
\end{equation}
where $i$ is the inclination of the orbit and $\omega _p$ is the longitude of the ascending node. Note that the orbit is assumed to be nearly circular, which is valid for HD 189733b. The rotation itself is of the form
\begin{equation}
Q'' = Q' \cos 2\gamma _2
\end{equation}
\begin{equation}
U'' = Q' \sin 2\gamma  _2
\end{equation}

\textit{I} and \textit{V} are unaffected as before. Note that $U'$ plays no role in the second rotation since its value drops to zero during the first rotation and summation over the illuminated disk for a planet that is symmetric about its equator. This set of transformations yield the full orbit polarized phase curve $[I(\phi),Q''(\phi),U''(\phi),V(\phi)]$ for the planet. Our simple model does not account for transit, secondary eclipse or limb effects in the star and the planet, non-spherical planets, thermal emission from the planet and other higher order effects. Those will be considered in future efforts.

\subsection{Atmospheric Structure of HD 189733b}

\begin{figure}[h!]
  \centering
    \includegraphics[width=9cm]{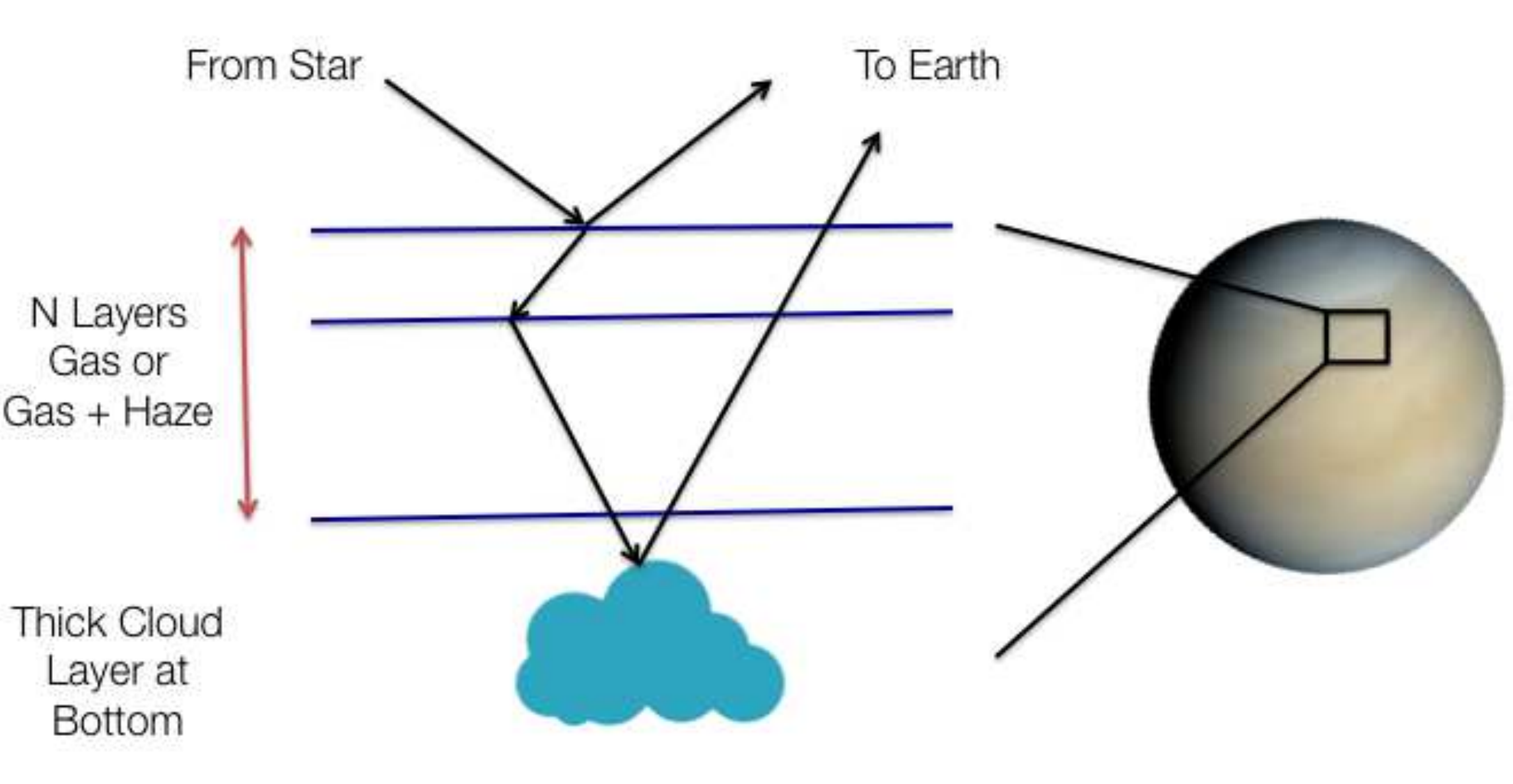}
      \caption{Simple schematic indicating the breakup of the illuminated disk into smaller regions, each one of which is represented by a stack of plane parallel atmospheric layers. There are N (N typically being 1 or 2) layers which can have either gas alone (Rayleigh scattering) or a mixture of gas and haze particles. This is underlain by a thick, reflective cloud layer at the bottom. Single and multiple scattering (as indicated by the black arrows) are calculated using the radiative tranfer model VLIDORT \citep{spurr2006vlidort}.}
       \label{fig:cartoon}
\end{figure}

Starting with the relatively complex Jupiter atmosphere of Z13, we construct simple atmospheric structures that may be plausible for HD 189733b. {The legacy atmospheric model from Z13 consists of 11 layers of gas and haze particles; gas is present in each layer while haze particles may or may not be present. The $12^{th}$ (deepest) layer being an optically thick reflective cloud layer with single scattering albedo 0.99. Gas is also present in this layer; however, the large optical depth of the cloud makes scattering by gas inconsequential within this layer. A simple schematic of this plane parallel atmosphere is provided in Figure \ref{fig:cartoon}. Since we do not require this level of vertical resolution with current observations, we will reduce the number of gas layers, \textit{N} to 1 or 2 depending on the case of interest. Should observations of sufficient quality become available, it is easy to add on more layers.} The cloud layer is underlain by a Lambertian surface of albedo zero to provide a boundary condition. However, the cloud layer is thick enough ($\tau _{cloud} = 50$) that changes to the albedo of this surface has no observable effect. 
The atmospheric composition of HD 189733b consists primarily of hydrogen and helium, with traces of methane, carbon dioxide and water. Since none of these gases have absorption lines or bands at $0.44 \mu m$, their contribution is primarily Rayleigh scattering. We take the typical depolarization ratio of 0.02 for hydrogen as representative of the atmosphere following \citet{stam2004using}. The scattering properties of the underlying cloud layer is described by a double Henyey-Greenstein (DHG) function, Equation 2 of Z13. {The DHG function is fully depolarizing}. For the sake of simplicity and due to the lack of better alternatives, we use the following values from Table 4 of Z13 for the parameters for the double Henyey-Greenstein scattering function, $f_1 = 0.8303$, $g_1 = 0.8311$ and $g_2=-0.3657$. {A summary of relevant atmospheric parameters is provided in Table \ref{tab:paramstable}.} The total column optical depth of the gaseous atmosphere (excluding the bottom cloud layer) is treated as a free parameter. However, we find that with total column optical depths of order one, a doubling or halving of the optical depth only results in changes of order 5-10\% in the intensity and degree of polarization for a pure Rayleigh scattering atmosphere.

Haze particles are either modeled as Mie spheres \citep{de1984expansion} or fractal aggregates, using the approximate method of \citet{tomasko2008model} designed for Titan hazes. The refractive index of the particles is fixed at $1.68+10^{-4}i$ \citep{jager2003steps}, representing a composition of silicate grains as hypothesized by \citet{pont2013prevalence}. Since the slope seen in transmission spectra is attributed to a scattering haze over several scale heights, we assume a well mixed atmosphere. Thus, the mixing fraction of haze particles will have no vertical variations in our models, unless the variation is the difference between the existence or absence of haze.
\begin{table}[h!]
  \begin{center}
    \caption{Summary of Parameters Used}
    \label{tab:paramstable}
    \begin{tabular}{ccc}
 \hline
 \hline
    Function & Parameter & Value  \\
 \hline
Wavelength & & 0.44 $\mu m$ \\
Cloud Ph. Fn. (DHG) & f1 & 0.8303\\
 & g1 & 0.8311 \\
 & g2 & -0.3657 \\
Haze Particles & Refractive Index & $1.68+10^{-4}i$ \\
& Radius (spherical) & $1\mu m$ \\
& Monomer radius (fractal) & $10 nm$ \\
& Monomers/particle  & 1000 \\
 \hline  

    \end{tabular}
  \end{center}
\end{table}
{For the following sections, we fix the wavelength of radiative transfer calculations at $0.44 \mu m$, which is the central wavelength of the B band in the visible range. It is relatively straightforward to change the wavelength to the UV or IR ranges, as long as the relevant scattering and  extinction cross-sections are available for all gases, haze and cloud particles. The planetary and orbital parameters for HD 189733b are taken as follows. The radius of the planet is 1.138$R_j$, semi-major axis is 0.03 AU and the eccentricity of the orbit is taken to be nearly zero (actual value is 0.0041) \citep{torres2008improved}. The inclination of the orbit can be either $86^{\circ}$ \citep{triaud2010spin,berdyugina2008first} or $94^{\circ}$\citep{berdyugina2011polarized}(we use $94^{\circ}$) and the longitude of the ascending node is $16^{\circ}$ \citep{berdyugina2008first}.} 
\section{Results and Discussion}
Polarimetric observations typically yield the two linear polarization parameters, Stokes \textit{Q} and \textit{U}, in addition to the intensity. {In reflected starlight, circular polarization over the northen and southern hemispheres will likely have comparable absolute values but opposite signs. The result is that integrated over the disk, the circular polarization values are very small.} The degree of circular polarization is at least 4-5 orders of magnitude smaller than the linear polarization, and cannot be measured with current technology for exoplanets. Thus, the reflected light from the atmosphere of an exoplanet is described fully by \textit{I}, \textit{Q} and \textit{U}. The total degree of polarization of reflected light is wholly determined by the nature of scattering in the planetary atmosphere, while the viewing geometries determine the distribution of polarization between the parameters \textit{Q} and \textit{U}. The broad atmospheric structure of HD 189733b is still a matter of active debate. Depending on the interpretation of transit spectra, cloudy or clear atmospheric scenarios cannot be ruled out \citep{crouzet2014water}. Thus, we will examine a few simple structures and their associated polarization signatures here.
\subsection{Semi-infinite Rayleigh Scattering Atmospheres}
\begin{figure}[h!]
  \centering
    \includegraphics[width=9cm]{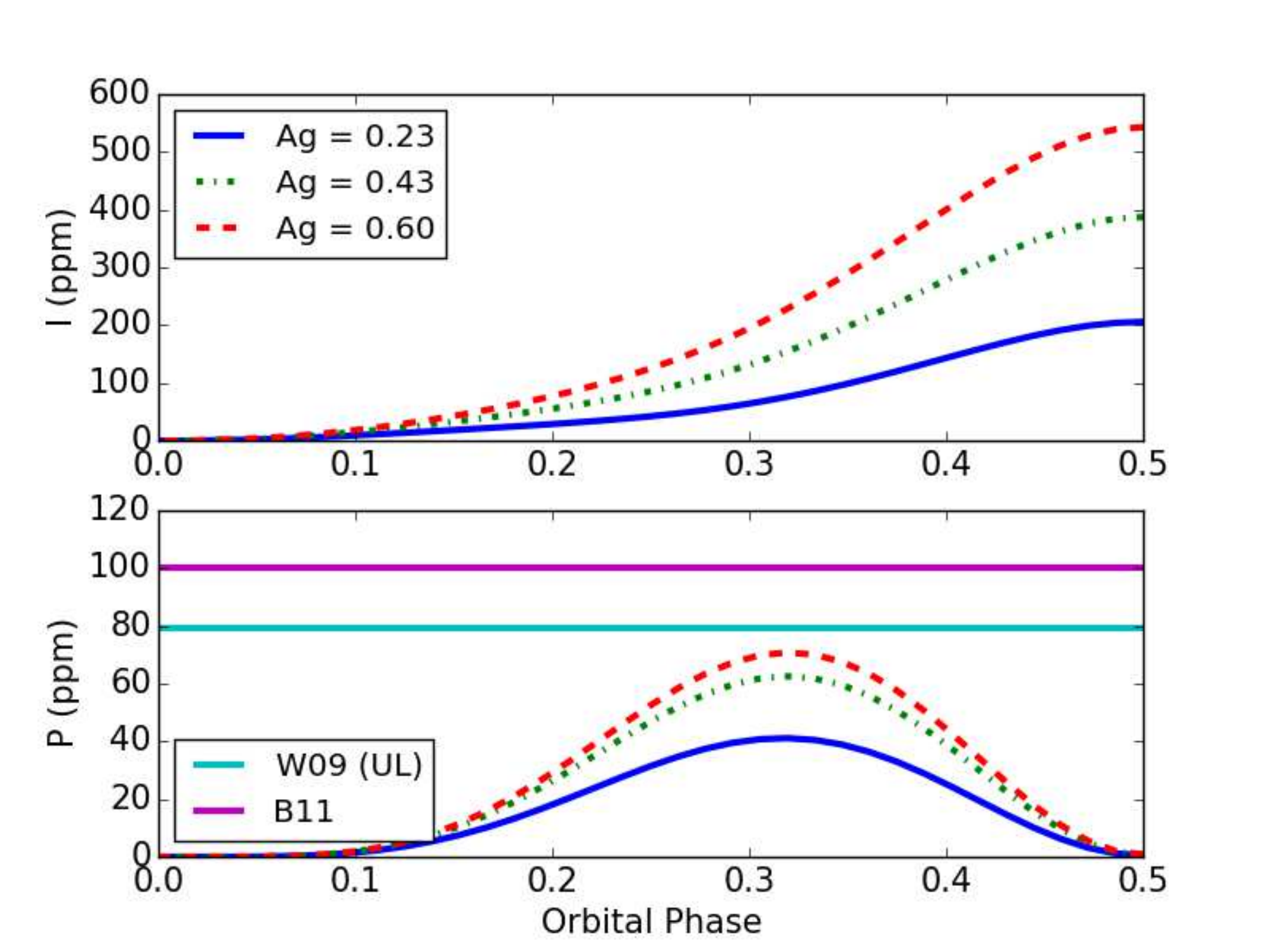}

      \caption{Variation in the degree of polarization for reflected light from HD189733b with changes in geometric albedo for a semi-infinite, purely Rayleigh scattering atmosphere. \textit{I} and \textit{P} are normalized to direct starlight. The B11 and W09 lines indicate the amplitude of observations from \citet{berdyugina2011polarized} and the upper limit for non-detection from \citet{wiktorowicz2009nondetection}. Orbital phase 0 is mid-transit and 0.5 is mid-eclipse.}
          \label{fig:albedo}
\end{figure}

 {The recent ideas of \citet{crouzet2014water,mccullough2014water} support a clear atmosphere for HD 189733b. Thus, we start with a very simple case: a thick, purely Rayleigh semi-infinite scattering atmosphere. The atmosphere has one single Rayleigh scattering layer with an optical depth of 1000, above a Lambertian surface of albedo 1.0}. We will refer to such atmospheres as a semi-infinite Rayleigh scattering atmospheres in the following discussions. We vary the  single scattering albedo to simulate the effect of changes in geometric albedo of the planet. The results are shown in Figure \ref{fig:albedo}, with reflected intensity and polariztion normalized to total direct starlight. Disk integrations are carried out at every $5^{\circ}$ in orbital position angle. Following the notation of \citep{berdyugina2011polarized}, orbital phase angle of $0^{\circ}$ (secondary-eclipse) corresponds to an orbital phase of 0.5, and an orbital phase angle of $180^{\circ}$ (transit) corresponds to orbital phase of 0.0.  We also overplot the detected magnitude of polarization from \citet{berdyugina2011polarized} and the upper limit from the non-detection from \citet{wiktorowicz2009nondetection}, while noting that the central wavelength of the filters used in these observations do not exactly coincide with our model wavelength of $0.44 \mu m$. We cannot explain the large value of \citet{berdyugina2011polarized} using this atmospheric structure, under the assumption that the polarization is due to reflection from the planetary atmosphere. {However, we note that the large value of observed polarization still points towards a highly reflective Rayleigh-like atmosphere, since any other type of scattering particle will reduce the degree of polarization.} 

\begin{figure}[h!]
  \centering
    \includegraphics[width=9cm]{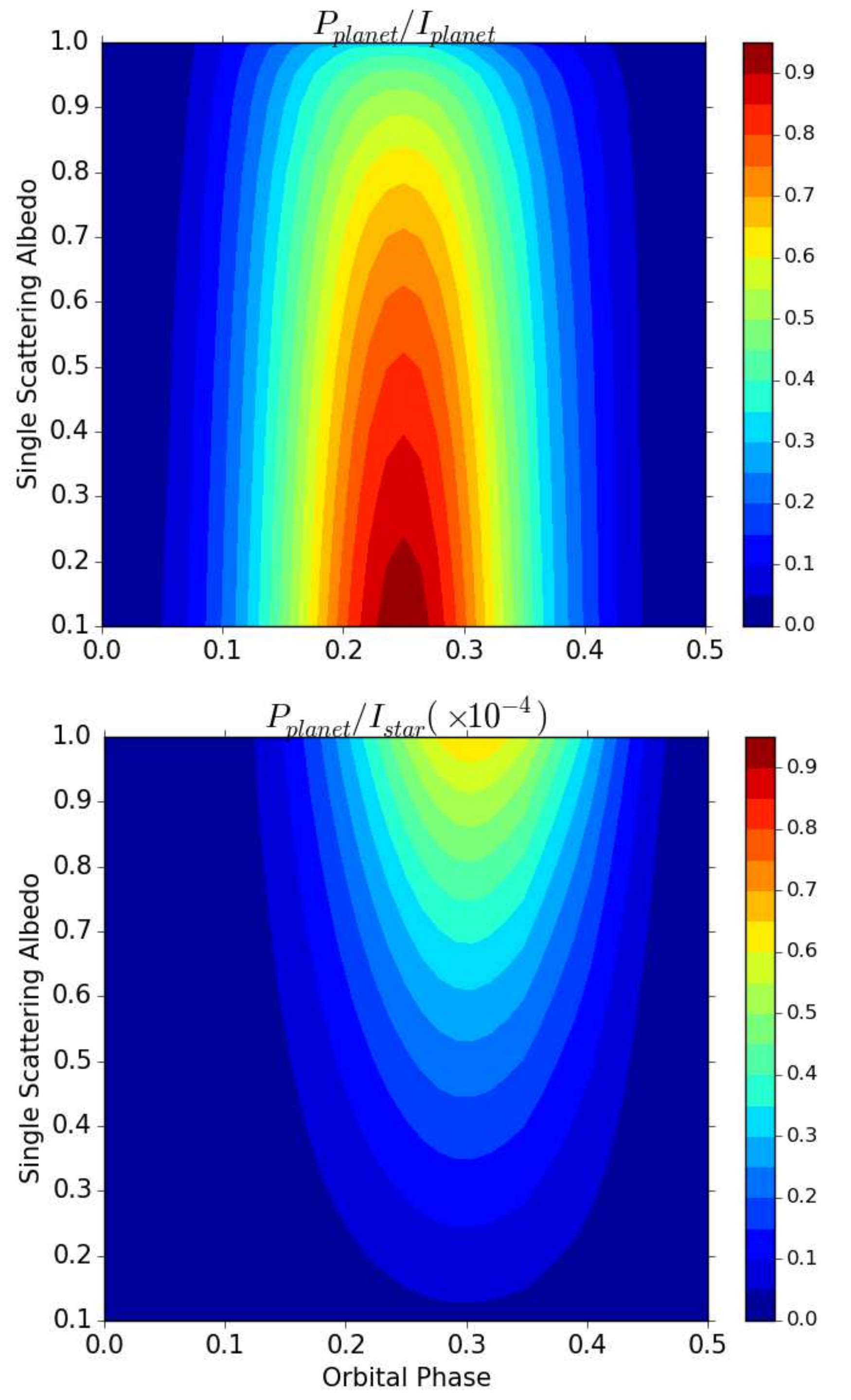}

      \caption{Variation in the degree of polarization as a function of single scattering albedo and orbital phase for a semi-infinite Rayleigh scattering atmosphere normalized to reflected light from the planet (top) and direct starlight (bottom). For the former, the highest degree of polarization occurs at low albedo, while for the latter (which is the observable quantity), it occurs at high albedo.}
          \label{fig:contours}
\end{figure}

{For a purely Rayleigh scattering atmosphere, the degree of polarization depends only on the single scattering albedo. The single scattering albedo accounts for the presence of absorbing gases in that layer. Since each photon is scattered around till it is either absorbed or leaves the atmosphere, layers with high albedo have multiple scattering that randomizes the plane of polarization and reduces the observed degree of polarization at the top of the atmosphere. One might be tempted to infer, therefore, that low albedos are preferable to reduce multiple scattering and have larger polarization signals. However, as the albedo of the atmosphere is lowered, the planet becomes dimmer with respect to the star. Consequently, the maximum degree of polarization in the star-planet flux becomes lesser. These two competing albedo effects give rise to different behaviors depending on whether the polarization is normalized to the intensity of the star or the reflected intensity of the planet as seen in Figure \ref{fig:contours}.}

For the following cases, we will hold the geometric albedo constant, unless otherwise stated, and vary other atmospheric parameters. We use a value of 0.23, which is the geometric mean of the observed albedos spanning the B-band ($\sim 390-480 $ nm) as reported in Table 1 of \citet{evans2013deep}.

\subsection{Semi-infinite Hazy Atmospheres}
\begin{figure}[h!]
  \centering
    \includegraphics[width=9cm]{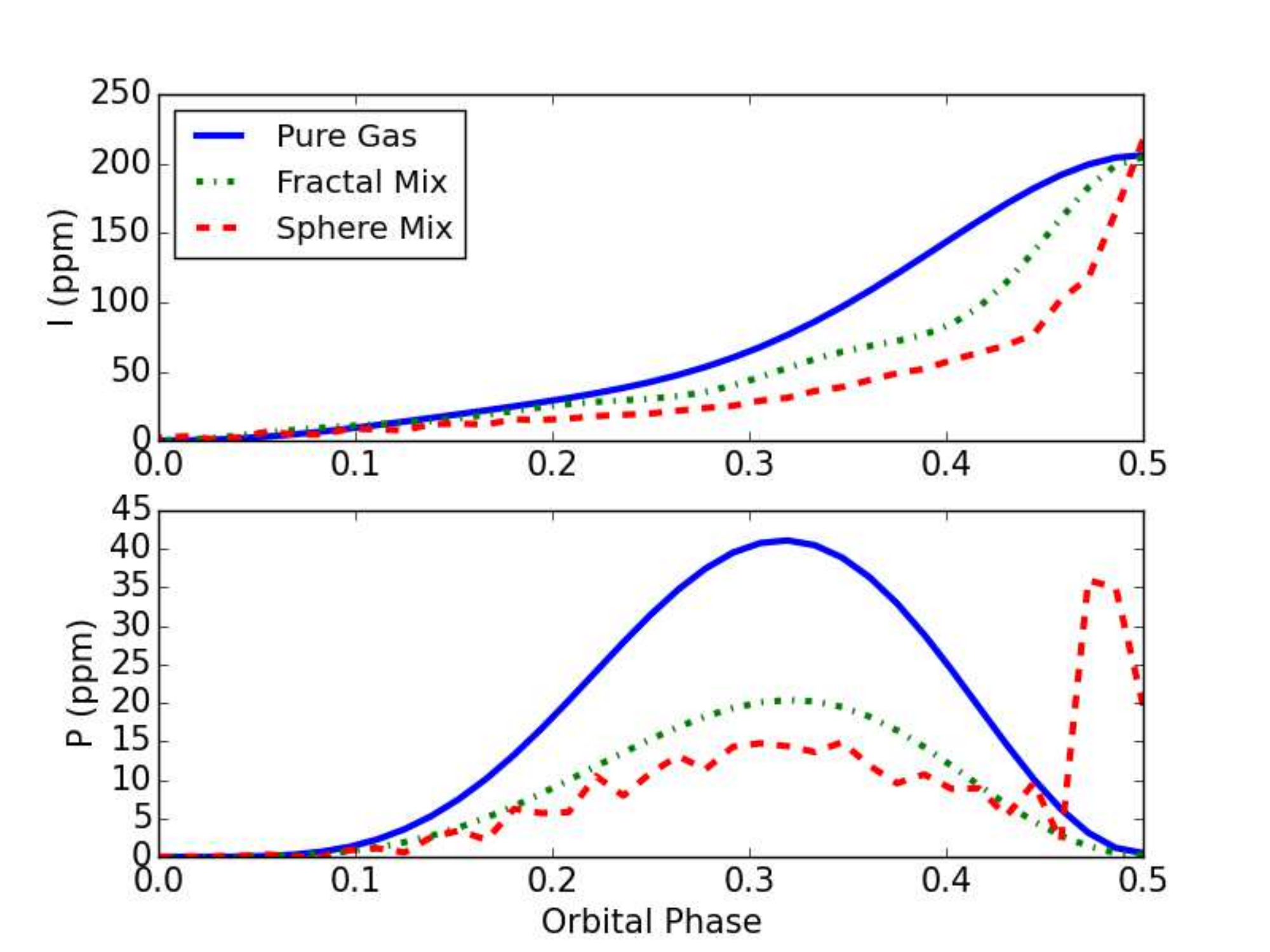}
      \caption{Variation in the degree of polarization from reflected light HD189733b system with a semi-infinite pure gas and hazy atmospheres. The particle properties are listed in Table \ref{tab:paramstable}. The geometric albedo of the planet is forced to remain close to 0.23. B11 and W09 lines indicate the amplitude of observations from \citet{berdyugina2011polarized} and \citet{wiktorowicz2009nondetection}.
          \label{fig:degpol}
}
\end{figure}

{ Based on the interpretation of \citet{pont2013prevalence} and others, the atmosphere of HD 189733b consists of a well-mixed Rayleigh scattering haze over several scale heights.} To model this, we introduce two types of scattering particles into the atmosphere: spherical particles of size $1 \mu m$ and fractal particles of effective size $\sim 0.1 \mu m$, composed of $1000$ spherical monomers of size $\sim 10 nm$. These are similar in shape to fractal particles used in Z13, but their refractive index is that of silicates, $1.68+10^{-4}i$. As in Section 3.2, a single gas+haze layer with an optical depth of 1000 makes up the atmosphere, with an underlying cloud layer. These particles are added such that they contribute to 50\% of the optical depth at each atmospheric layer, while the geometric albedo is held constant close to 0.23. This is achieved by setting the single scattering albedo to 0.71 in the Rayleigh case, 0.54 in the Mie case and 0.84 in the fractal case. The resulting curves are shown in Figure \ref{fig:degpol}.\footnote{ {The jagged appearance of the Mie curve is due to disk integrations being carried out at every $5^{\circ}$ in orbital phase angle. Smooth curves can be obtained by carrying out integrations at every $1^{\circ}$ (Appendix, Figure 14), but the increased computational time does not yield any fundamentally new features.}} The highest polarization is always produced by a non-absorbing, purely Rayleigh scattering atmosphere. The introduction of any particle that deviates from this regime reduces the polarization. {Polarization is non-zero at orbital phase 0.5 since the planet is in an orbit whose inclination is not $90^{\circ}$. Therefore, at this orbital phase the phase angle is $\sim 4^{\circ}$, while polarization is zero for a phase angle of $0^{\circ}$.}  
 
 A simple reason to explain this effect is that moving from the Rayleigh to Mie regime reduces reflection at quadrature angles and increases preferential forward scattering. Since the polarization peak occurs near quadrature, and there is lower reflection at this point, the total degree of polarization invariably decreases. {The fractal particles are characterized by their Mie particle-like intensity curve, which comes from their large effective radius and Rayleigh-like polarization curve, which is due to the small size of individual monomers that make up the aggregate. The Mie-particle haze can be distinguished by a rainbow close to secondary eclipse. Thus, for a given albedo, using a combination of intensity and polarization measurements, it should be possible to determine whether a haze is present, and what type of particles might be present in it.{Recent work has begun to place constraints on scattering particle properties \citep{munoz2015probing}}. Increasing effective haze particle size decreases the degree of polarization observed. However, it will be tricky to characterize the size of haze particles from the degree of polarization alone without extremely high resolution polarimetric observations ($\Delta P \sim$ few ppm).}

\subsection{Thin Atmospheres Above Cloud Decks}
\begin{figure}[h!]
  \centering
    \includegraphics[width=9cm]{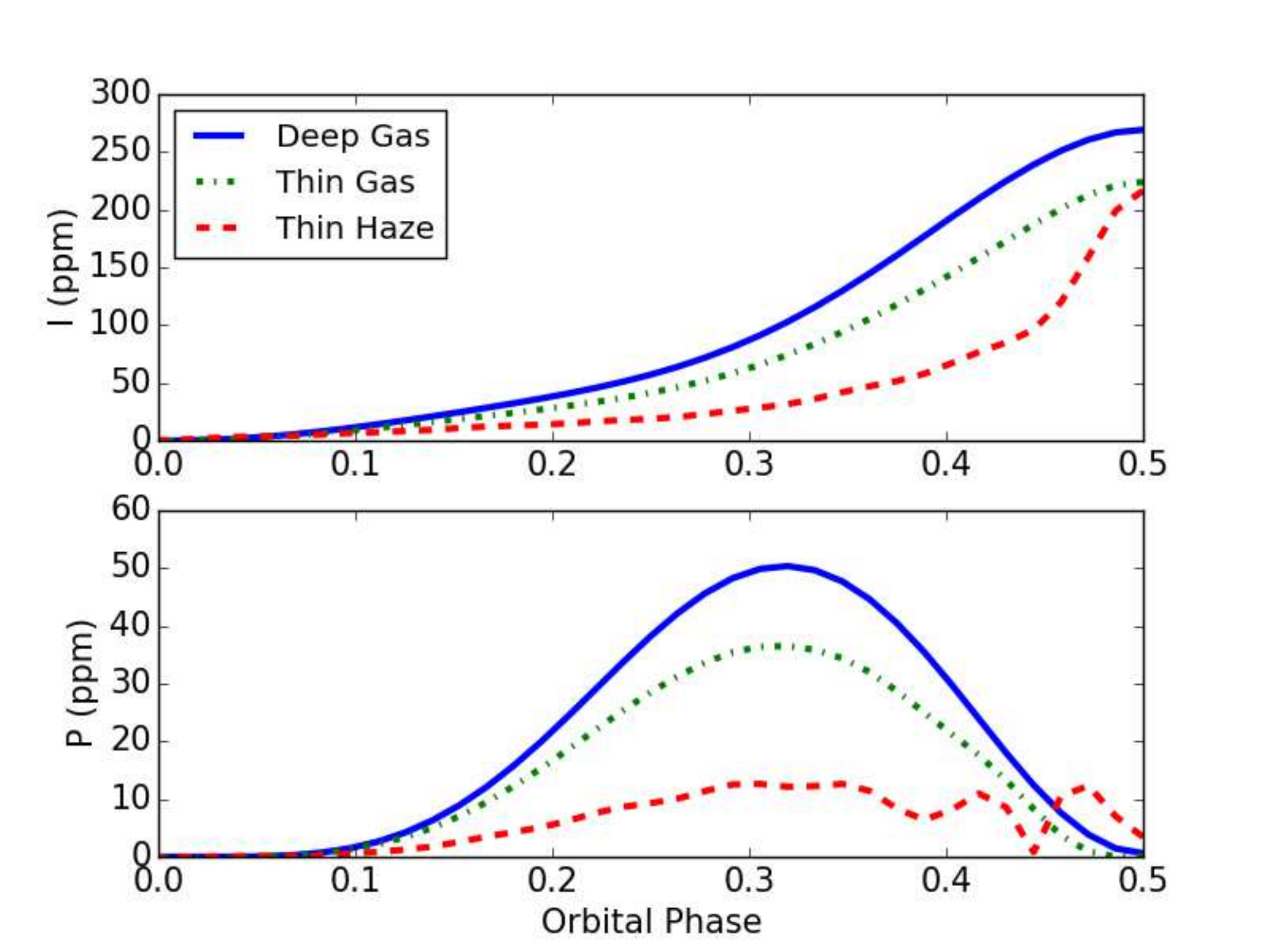}

      \caption{Variation in reflected intensity and the degree of polarization  for different atmospheric structures of HD 189733b. The intensity curves for a semi-infinite Rayleigh atmosphere (deep gas), thin, clear gas atmosphere (thin gas) and a hazy atmosphere with spherical particles (thin haze) on top of a cloud layer. The haze and cloud properties are mentioned in Table \ref{tab:paramstable}.}
          \label{fig:structures}
\end{figure}

 { \cite{berdyugina2011polarized} proposed an atmospheric structure with a thin gas or haze layer on top of a semi-infinite cloud or condensate deck. Since the nature of the cloud or haze layer remains fairly unconstrained in this picture, we create a structure with two layers. In the first case, the top layer is pure gas with an optical depth of 1.0, and single scattering albedo of 0.7. In the second case the top layer has spherical haze particles and gas, each component contributing to $50\%$ of the optical depth with a total optical depth of 1.0 and single scattering albedo of 0.5. The bottom layer is a cloud with optical depth of 1000 and albedo of 0.7, while the cloud scattering properties are represented by the DHG function mentioned in Table \ref{tab:paramstable}.} Scattering by the cloud produces no net polarization. This is to simulate the effects of scattering by very large cloud particles, of millimeter size. In all cases, the geometric albedo of the planet is maintained close to 0.23. The results are shown in Figure \ref{fig:structures}, we compare these cases to a semi-infinite Rayleigh scattering atmosphere since that is the basic structure that we must distinguish from. Since the geometric albedo is constrained to be the same in all cases, changes in observed intensity are very minor. {Thin polarizing layers produce a lower degree of polarization, but the shapes of the curves are the same as those of thick atmospheres of similar composition in the previous section. There is no particular advantage to using polarimetry in this case. This information can also be acquired from a different observational technique, such as transit photometry.}
\subsection{Inhomogeneous Atmospheres}
Thus far we have considered homogeneous atmospheres, in both the vertical and horizontal directions, which are idealized cases. We treat one case of horizontal inhomogeneity, where one hemisphere is covered by a haze and the other hemisphere is clear. Such scenarios are of particular interest, since haze and cloud formation process often produce patchy, inhomogeneous regions as seen in the Solar System planets and brown dwarfs. A recent study of the exoplanet Kepler 7b indicates the presence of spatial inhomogeneity where one hemisphere of the planet is more reflective than the other \citep{demory2013inference,hu2015semi}, possibly indicating that one hemisphere is covered by patchy clouds while the other is clear. 
 
\begin{figure}[h!]
  \centering
    \includegraphics[width=9cm]{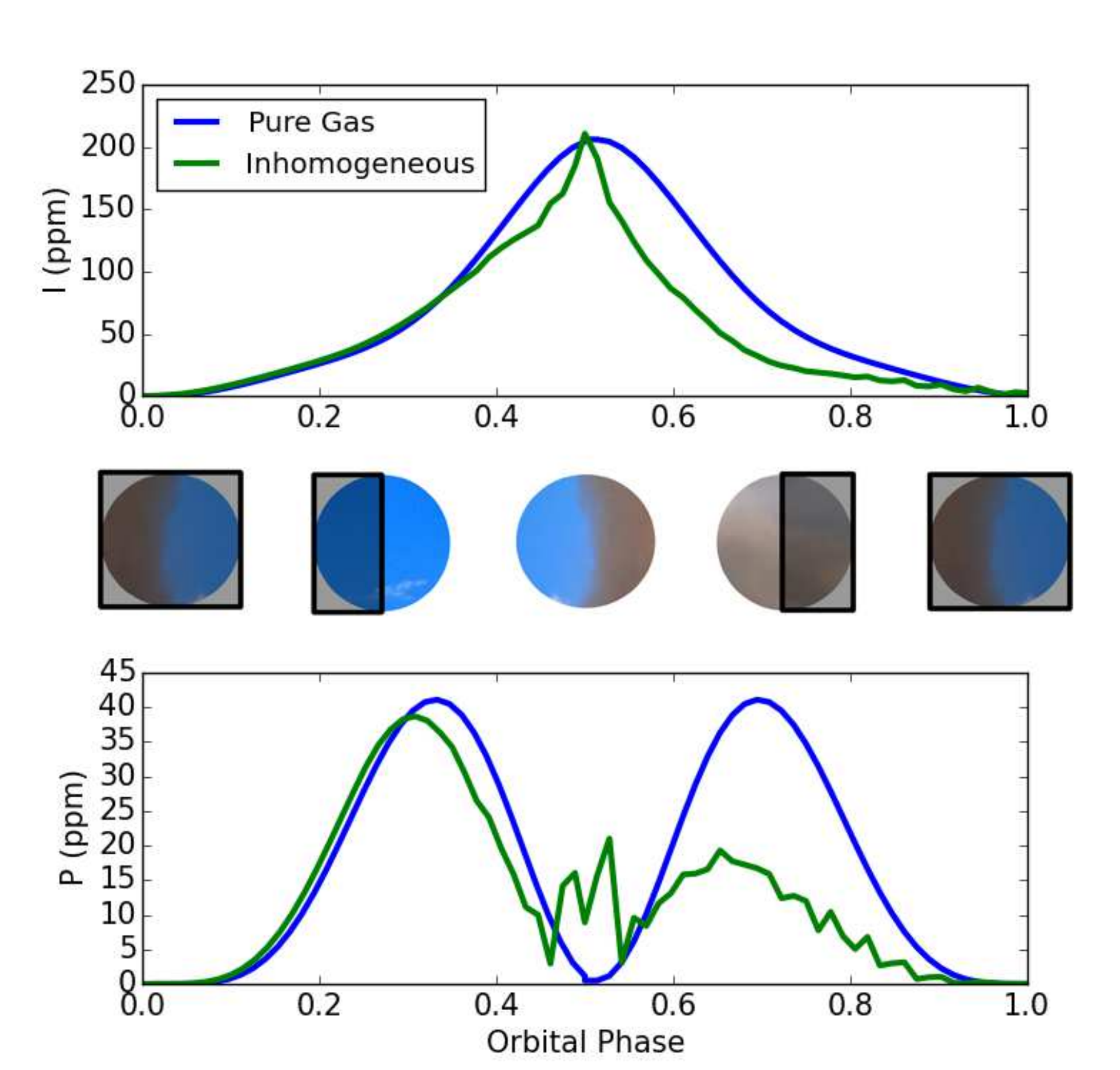}

      \caption{Variation in reflected intensity and the degree of polarization as a planet with an inhomogeneous atmosphere completes one orbit compared to a homogeneous, Rayleigh scattering planet. The spheres on top show the planet as seen from Earth at the phases indicated on the abscissa. The dark blue regions are pure Rayleigh scattering, and the greyish regions contain haze. The portion covered by the box indicates the night side of the planet. }
          \label{fig:phases1}
\end{figure}

Here we assume that one hemisphere has a semi-infinite Rayleigh atmosphere (as in Section 3.1) and the other has a semi-infinite hazy atmosphere with spherical particles(as in Section 3.2). The hazy hemisphere has an effective geometric albedo of 0.19, to simulate the effect of greater scattering and absorption, while the geometric albedo of the Rayleigh hemisphere is maintained at 0.23. The hazy hemisphere covers the western part of the planet, lying half over the dayside and half over the nightside, as seen from Earth at secondary eclipse (Figure \ref{fig:phases1}). The peak of the reflected intensity is now just before eclipse. Note that the contrast between the homogeneous and inhomogeneous cases is exaggerated in the degree of polarization at quadrature as compared to the reflected intensity. 

Numerically, we create two different atmospheric structures. All longitudes west of the substellar point (which lies at the longitude equal to the scattering angle, $\alpha$) correspond to the clear structure, eastward are hazy. Thus far, we have used $\alpha$ as defined by Equation 4, which only yields non-negative values. We can get away with only positive $\alpha$ for a homogeneous planet because of longitudinal symmetry. For an inhomogeneous planet, we must have negative $\alpha$ values between $\phi=[\pi,2\pi]$ to ensure that the correct scattering angles are used. {Inhomogeneous atmospheres have been modeled by \cite{karalidi2012modeled,karalidi2013flux}, by calculating the brightness of homogeneous planets and creating an inhomogeneous planet from their area-weighted averages. One advantage of this method is that we do not need to repeat calculations for different homogeneous planets before arriving at the inhomogeneous case.} 

\subsection{Dependence on Orbital Parameters}
\begin{figure}[h!]
  \centering
    \includegraphics[width=7cm]{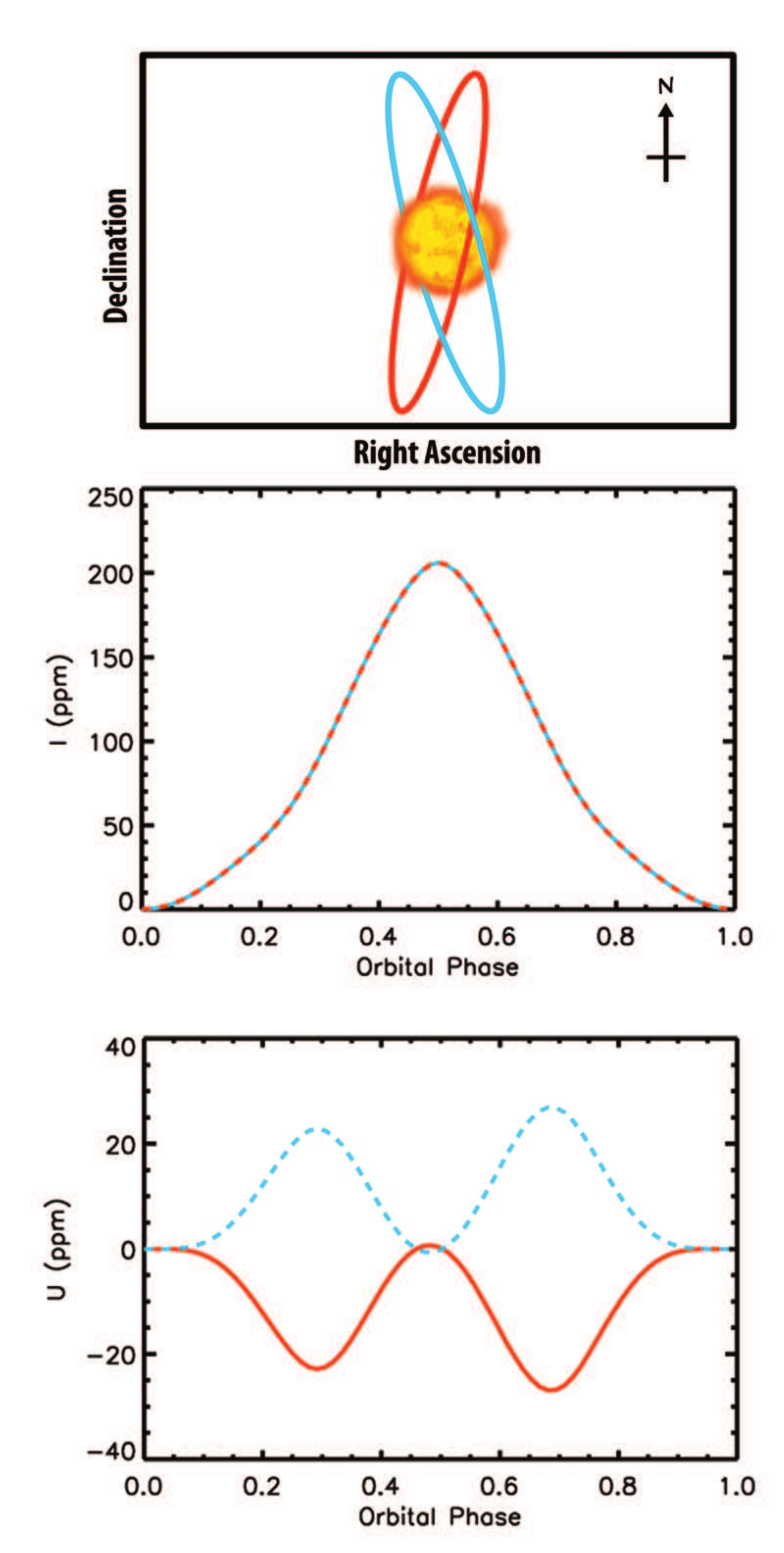}

      \caption{The top panel shows a cartoon of two orbits of inclination close to 90 degrees and longitude of the ascending node 16 degrees (red, solid line) and -16 degrees (blue, dashed line) for the HD189733b system as seen from Earth. (Figure is approximate, not to scale, angles are not accurately depicted). The arrow indicates the sense of motion of the planet in the orbit and upwards is North in the sky plane of the Earth. These orbits are indistinguishable from photometry alone, but can be separated using polarimetry. The sign of Stokes parameter \textit{U} changes, while intensity is invariant for this pair of orbits.}
          \label{fig:orbits}
\end{figure}

 The range of observed phase angles for one orbit of the exoplanet around the star is set entirely by its inclination. For instance, an inclination of $0^{\circ}$, allows only a constant phase angle of $90^{\circ}$, while an inclination of $90^{\circ}$ allows the full range from $0-180^{\circ}$. Intermediate values of inclination allow smaller ranges of phase angles to be observed. Since the inclination can usually be inferred from the transit light curve, we do not consider it a free parameter. However, the longitude of the ascending node cannot always be pin pointed from photometric light curves alone. Figure \ref{fig:orbits} shows an example of two possible transiting orbit candidates for an exoplanet which have the same inclination, but longitudes of the ascending node are of opposite sign albeit same magnitude.  The first panel shows the photometric light curve, which is identical for both orbits and cannot be used to distinguish them, while the polarimetric curve, \textit{U}, clearly shows a change in sign.
\section{Conclusions}
In this paper, we describe a multiple scattering radiative transfer model capable of generating polarized phase curves for reflected light from a range of atmospheric structures.{ In general, we find that our multiple scattering model cannot produce polarization high enough to match the observations of \cite{berdyugina2011polarized}, agreeing with the findings of \cite{lucas2009planetpol}. We also find that clear and hazy atmospheres have observable differences in polarized light. In combination with full orbit reflected intensity phase curves, it might be possible to even distinguish if the haze particles are spheres or aggregates. Furthermore, we also find that spherical haze particles with the refractive of silicate have a rainbow, and corresponding peak in polarization, close to secondary eclipse. In addition, we examine cases where a thin atmosphere is underlain by a semi-infinite cloud layer, and find that they are distinguishable from semi-infinite clear gas atmospheres. The semi-infinite Rayleigh scattering cases were used to put an upper limit on the albedo of HD 189733b in the visible in a companion paper \citep{wiktorowicz2015ground}}  

In light of growing interest in the exoplanetary community on classifying exoplanetary atmospheres as cloudy \citep{kreidberg2014clouds,knutson2014featureless} or clear \citep{fraine2014water}, polarimetry has great potential as an observational tool. The inferences of clouds through the transit observations use the absence of features in the spectra to postulate the presence of clouds. The inherent assumption here is that a thick cloud layer must cover a dominant fraction of the planet's atmosphere so as to mask absorption features. Note that clouds, at least those seen within the solar system are never uniformly thick or homogeneous (with the possible exception of Venus). There is no reason to expect that exoplanetary clouds will be any different. Thus, even exoplanets which show absorption features in their transit spectra might still admit patchy clouds in their atmospheres. The detection of patchy clouds is at the limit of current observational capabilities using photometric intensity alone, and must be indirectly inferred \citep{demory2013inference}. 

We show in this paper that contrasts between clear skies and fully or patchy clouds are significant in polarized light even when the reflected light intensities cannot be differentiated. The locations of hazes and clouds, combined with temperature profiles can be used to infer the composition of the condensates based on their condensation temperatures. {While intensity phase curves may yield information about the size of the scattering particle, polarized curves also give information about the refractive index depending on the position of the rainbow, allowing for additional constraints on chemical composition.} The size of cloud particles is indicative of the strength of the updrafts necessary to buoy them, among other factors (see \cite{reutter2009aerosol} for example,) and can provide constraints on the dynamics of exoplanetary atmospheres. The closeness of hot Jupiters to their stars, and the resulting interactions with stellar magnetospheres can influence the chemistry of the atmosphere. In the solar system, it is thought that the magnetosphere of Jupiter plays a key role in the creation of fractal aggregate hazes near the polar regions \citep{wong2003benzene}.

 Better constraints on the scattering properties of atmospheric particles and condensates will allow for the understanding of their formation mechanisms, which are linked to the circulation of the atmosphere itself. Though our model uses overly simplified atmospheric structures in its present form, future work will include spatial variations in atmospheric composition and structure in a more rigorous fashion. One possible extension might be to generate clouds and hazes through a 3D general circulation model and perform vector radiative transfer on the resulting atmospheric structures. As polarimetric observations converge on acceptable values for HD189733b, and new observations become available for other exoplanets, our model can be used in a retrieval framework to constrain atmospheric scattering properties and orbital elements. 

\section{Acknowledgements}
We thank David Crisp and Renyu Hu for several insightful comments. The paper also greatly benefited from the anonymous referee's reviews. This research was supported in part by the Presiden't and Director's Fund at Caltech and by the NAI Virtual Planetary Laboratory grant from the University of Washington to the Jet Propulsion Laboratory and California Institute of Technology. Part of the research described here was carried out at the Jet Propulsion Laboratory, California Institute of Technology, under a contract with the National Aeronautics and Space Administration. X.Z. was supported by the Bisgrove Scholar Program at the University of Arizona.
\newpage
\bibliographystyle{apj}

\section*{Appendix: Computational Considerations}
Here we show some the effects of varying different computational parameters in our model. The atmospheres in this section are semi-infinite and have a geometric albedo close to 0.23. The Rayleigh atmosphere is as described in Section 3.1 and the hazy atmosphere with spherical particles as described in Section 3.2. 
\begin{figure}[h!]
  \centering
    \includegraphics[width=7cm]{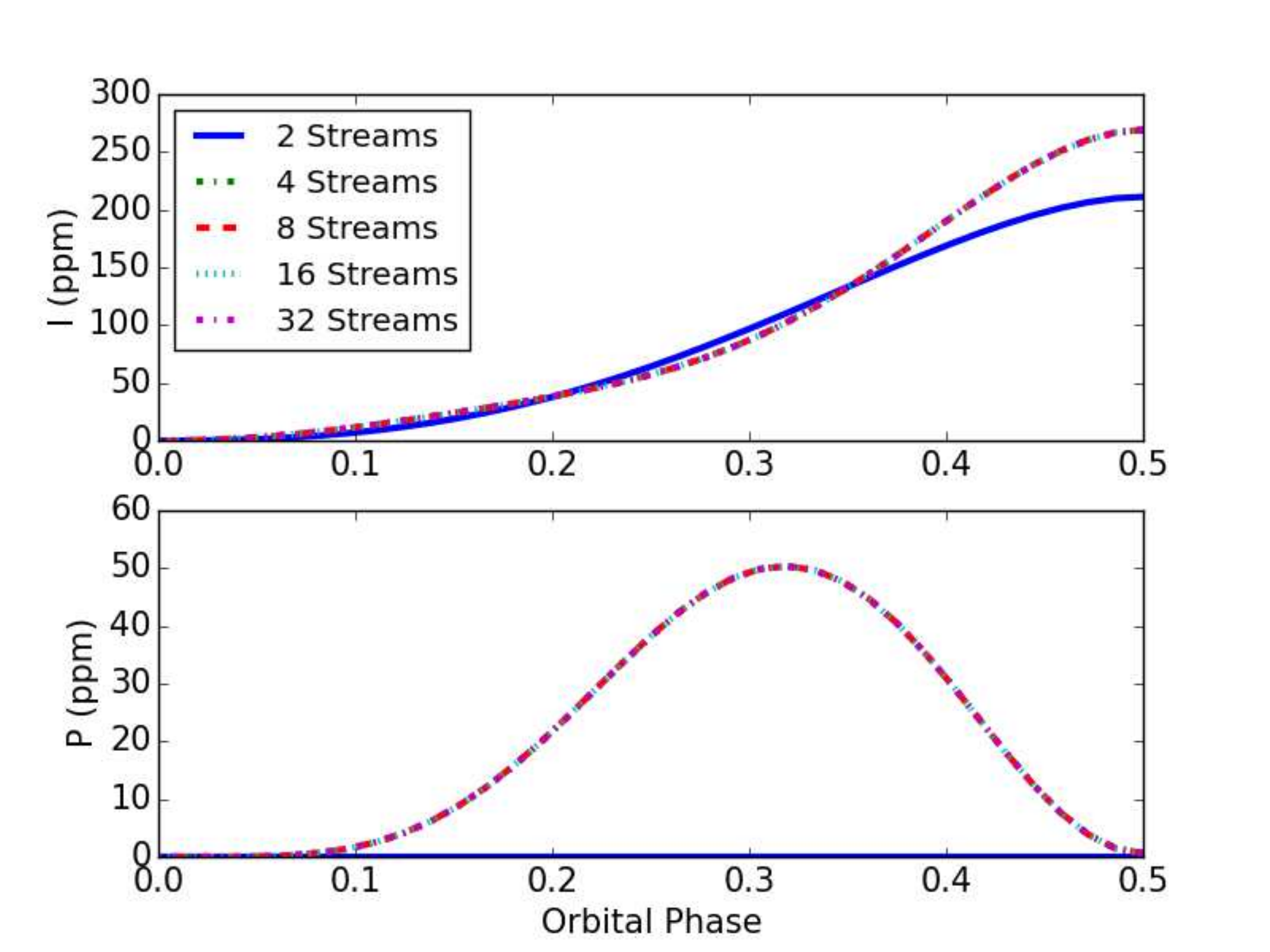}
      \caption{Model outputs for different number of computational streams in the RT model for a Rayleigh scattering atmosphere using 64 quadrature points for disk integration. 4 streams at least are necessary to produce polarization, but beyond that results are insensitive to change in streams. }
\end{figure}
\begin{figure}[h!]
  \centering
    \includegraphics[width=7cm]{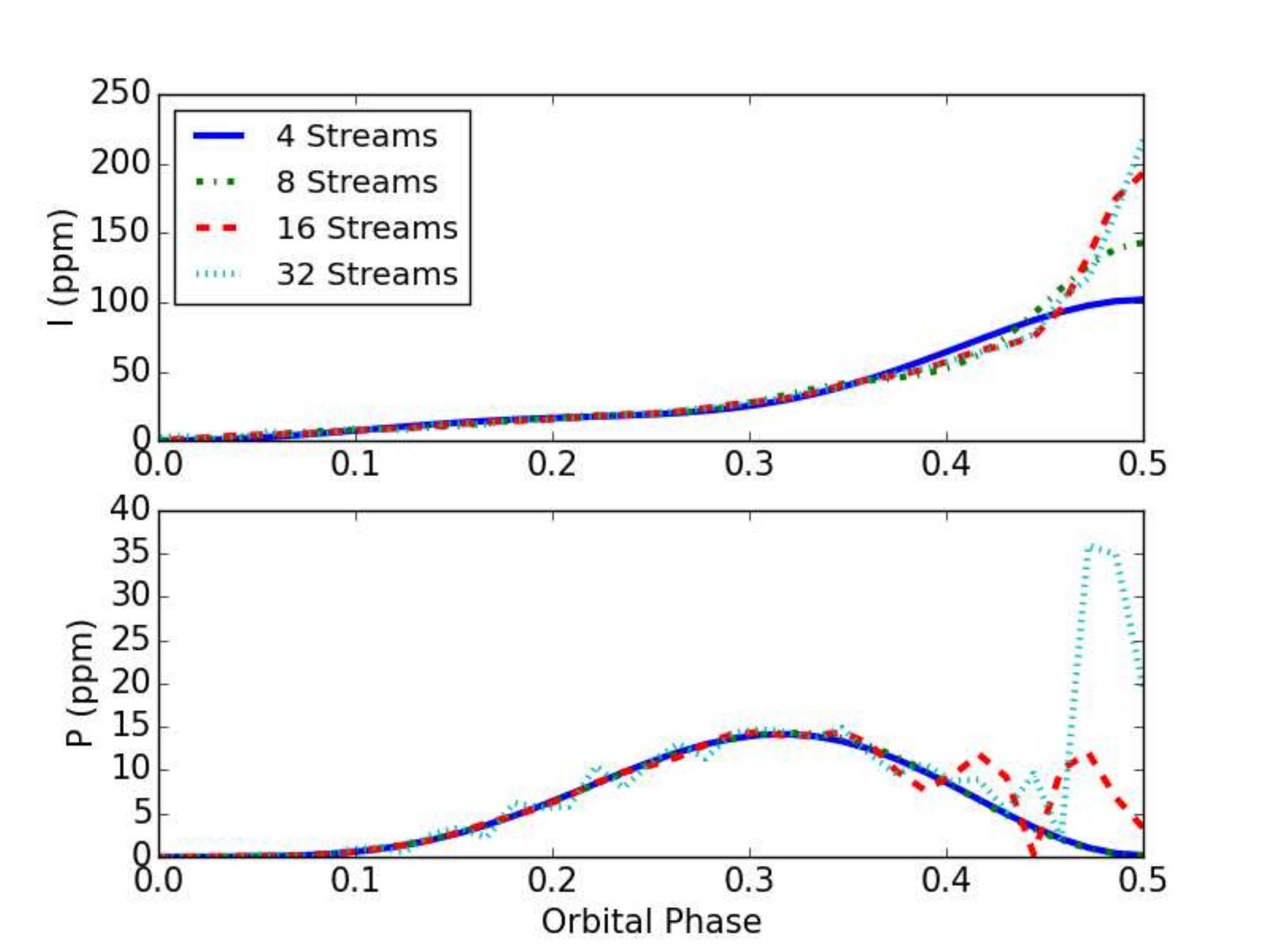}
      \caption{Model outputs for different number of computational streams in the RT model for a hazy scattering atmosphere using 64 quadrature points for disk integration. 16 streams at least are necessary to produce a rainbow. Hazy models in the paper are run with 32 streams and 256 quadrature points. The inhomogeneous planet uses 1024 quadrature points.}
\end{figure}
\begin{figure}[h!]
  \centering
    \includegraphics[width=7cm]{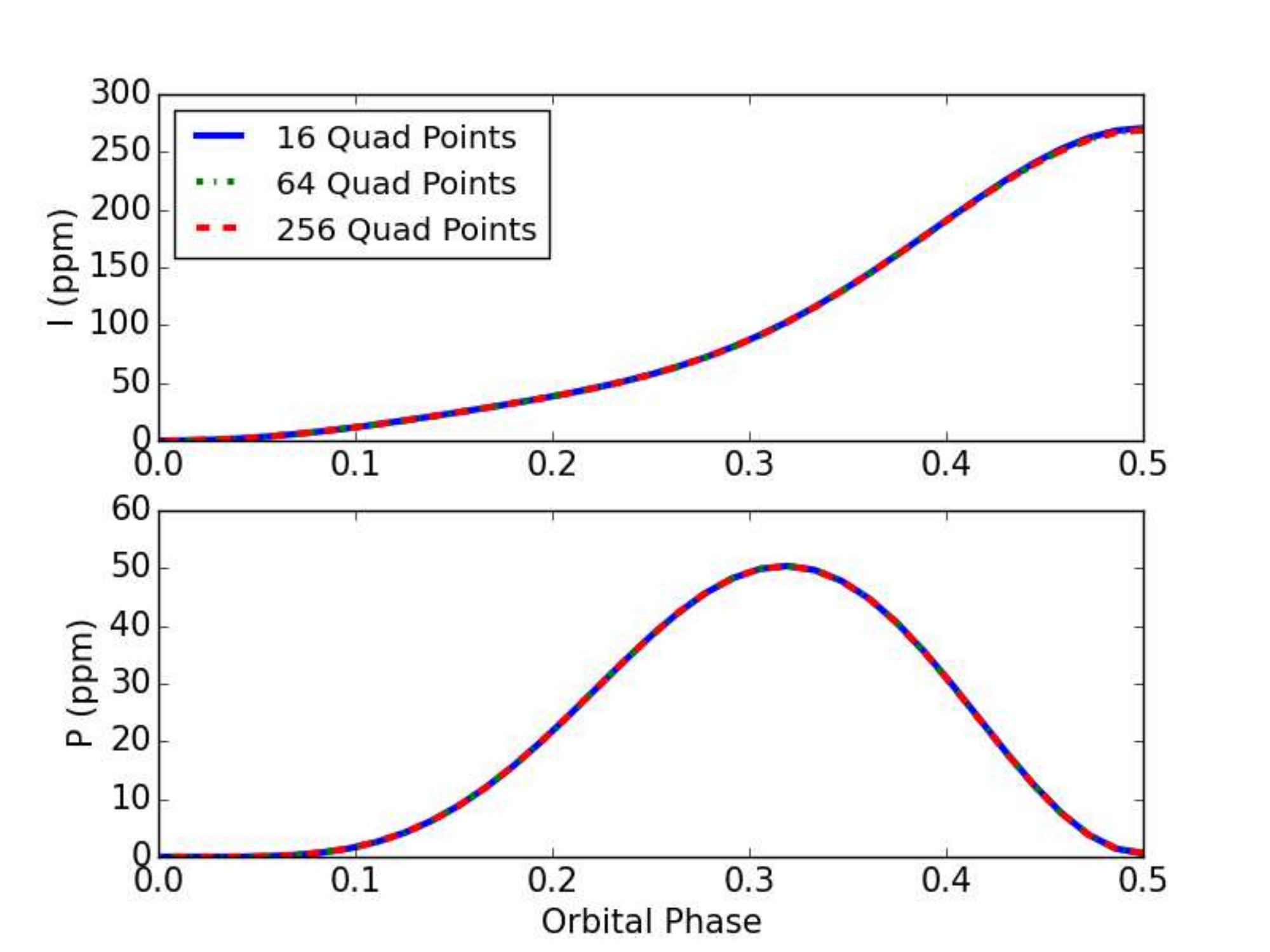}
      \caption{Model outputs for different number of quadrature points for a Rayleigh atmosphere with 8 computational streams for RT. Results are insensitive to the number of quadrature points.}
\end{figure}
\begin{figure}[h!]
  \centering
    \includegraphics[width=7cm]{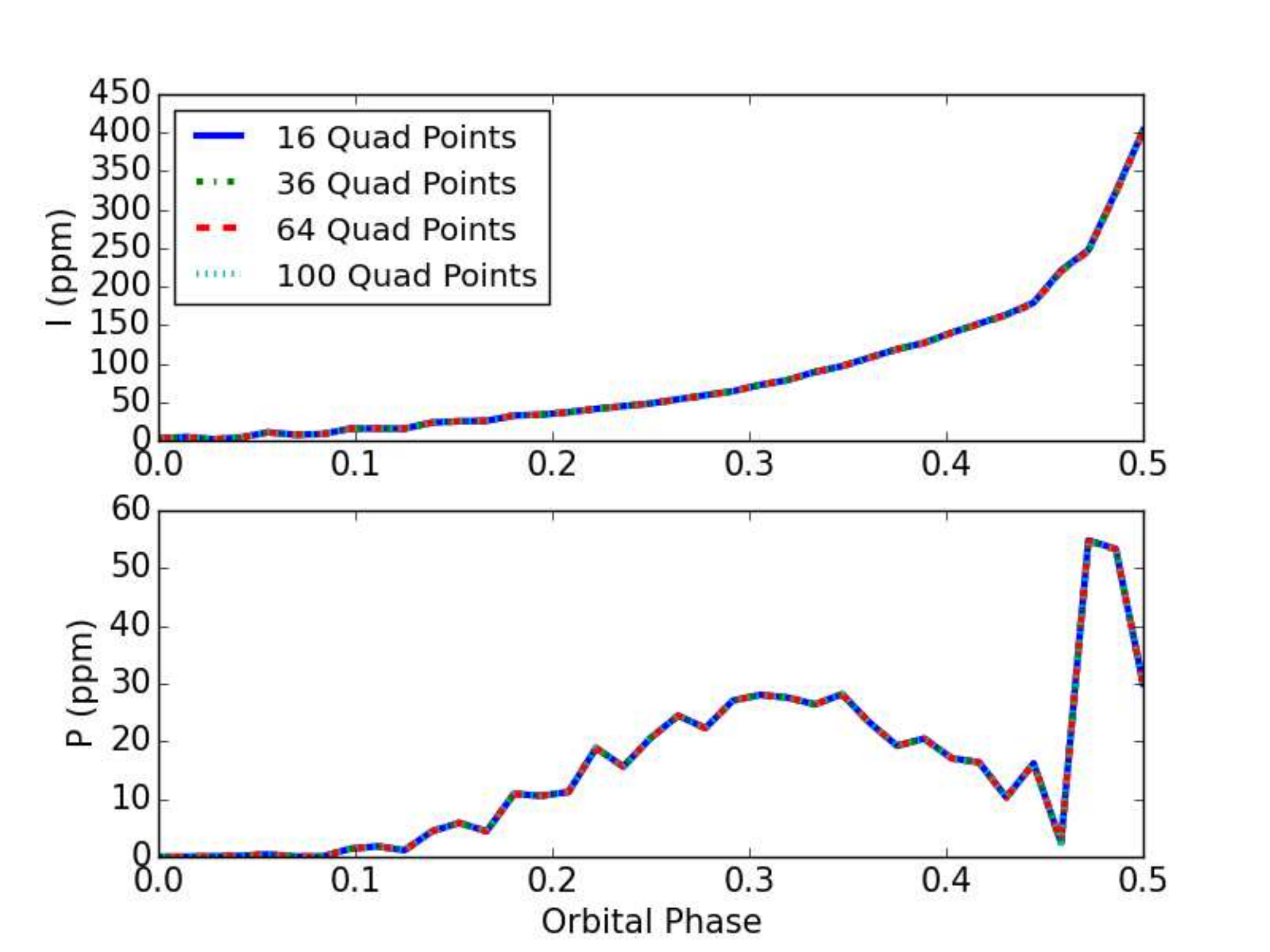}
      \caption{Model outputs for different number of quadrature points for a hazy atmosphere with 32 computational streams for RT. Results are sensitive to the number of quadrature points, but for the purposes of our discussion, the broad features are unchanged. Beyond 16 streams, the rainbow is always visible and the general shape of the curves remains the same in both \textit{I} and \textit{P}.}
\end{figure}
\begin{figure}[h!]
  \centering
    \includegraphics[width=7cm]{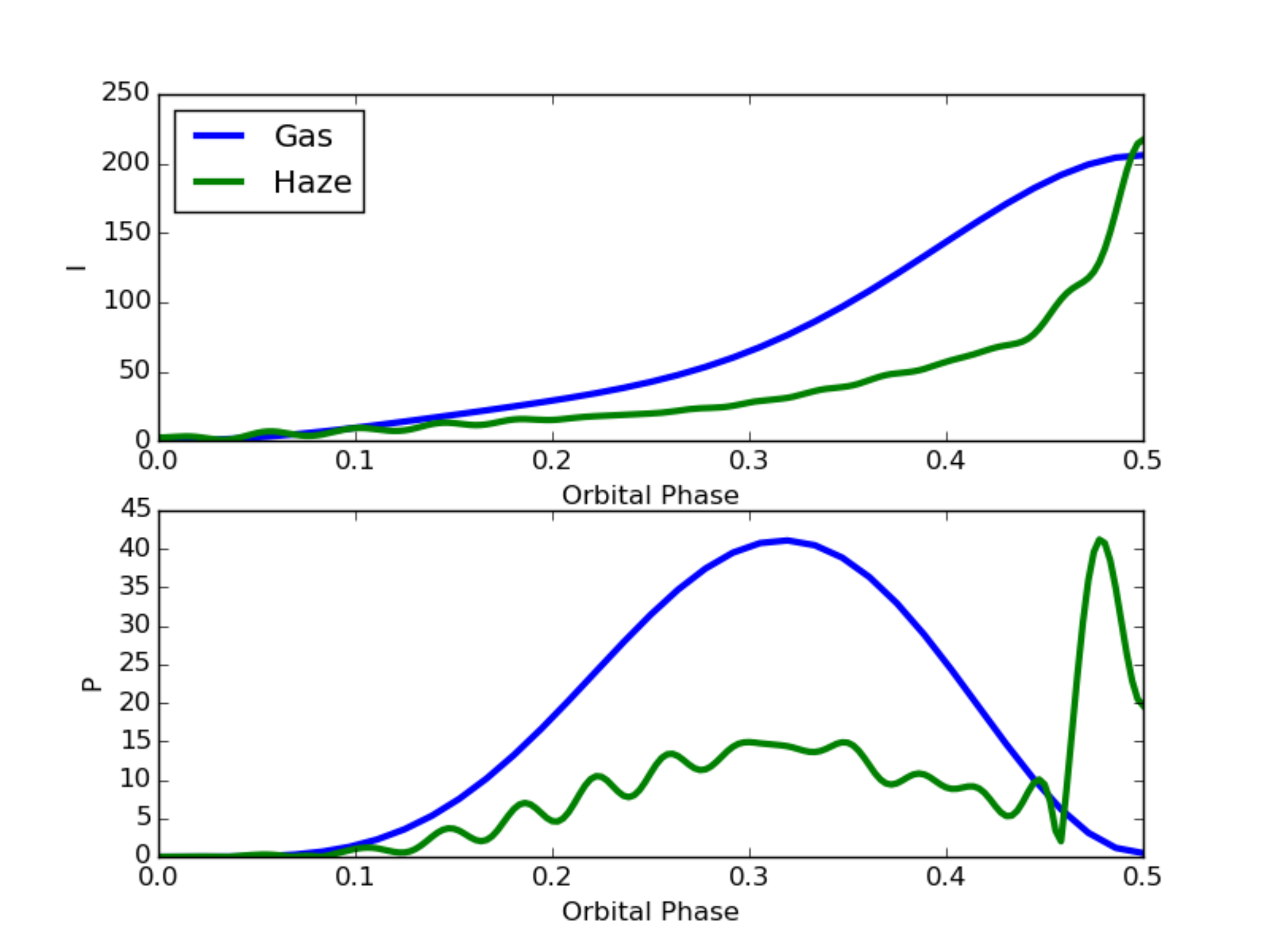}
      \caption{Model outputs for Rayleigh and hazy atmospheres with 32 streams and 256 quadrature points with a resolution of $1^{\circ}$ in orbital phase angle. Orbital phase from 0-0.5 is $180^{\circ}$, and typical resolution for all runs in this paper is $5^{\circ}$. Note the smooth waviness of the hazy curve. We continue to use $5^{\circ}$ since it does not miss any major features and has a significantly lower computational cost.}
\end{figure}
\end{document}